\documentclass[prd,twocolumn]{revtex4}
\usepackage[utf8]{inputenc}
\usepackage{graphicx}
\usepackage{amsmath}
\usepackage{amssymb}
\usepackage{color}
\usepackage{algorithmicx}
\usepackage{algpseudocode}
\usepackage{xspace}
\usepackage{multirow}

\newcommand{\argmin}{\operatornamewithlimits{arg\ min}}
\newcommand{\niter}{N_{\rm iter}}
\newcommand{\nruns}{N_{\rm runs}}
\newcommand{\snr}{{\rm SNR}}
\newcommand{\fitalgoname}{\texttt{SHAPES}\xspace}
\newcommand{\altalgows}{\texttt{WaveShrink}\xspace}
\newcommand{\altalgors}{\texttt{R:sm.spl}\xspace}

\begin{document}
\title{Adaptive spline fitting with particle swarm optimization}
\author{Soumya D. Mohanty}
\email{soumya.mohanty@utrgv.edu}
\address{Department of Physics and Astronomy, The University of Texas Rio Grande Valley, One West University Blvd., Brownsville, TX 78520, USA}
\date{March 2019}
\author{Ethan Fahnestock}
\email{efahnest@u.rochester.edu}
\address{Department of Physics and Astronomy, The University of Rochester, 500 Wilson Blvd., Rochester, NY 14627, USA}
\date{March 2019}
%%%%%%%%%%%%%%%%%%%%%%%%%%%%%%%%%%%%%%%%%%%%%%%%%%%%%%%%

\begin{abstract}
    In fitting data with a spline,  finding the optimal placement of knots  can significantly improve the quality of the fit. However, the challenging high-dimensional and non-convex optimization problem associated with completely free
    knot placement has been a major roadblock in using this approach.  We present a method that uses particle swarm optimization (PSO) combined with model selection to address this challenge. The problem of overfitting due to knot clustering that accompanies free knot placement is mitigated in this method by explicit regularization, resulting in a significantly improved performance on highly noisy data. The principal design choices available in the method are delineated and a statistically rigorous study of their effect on performance  is carried out using simulated data and a wide variety of benchmark functions. Our results demonstrate that PSO-based free knot placement leads to a viable and flexible adaptive spline fitting approach that allows the fitting of both smooth and non-smooth functions.
\end{abstract}
%%%%%%%%%%%%%%%%%  ARXIV PREPRINT %%%%%%%%%%%%%%%%%%%%%%
 \maketitle
%%%%%%%%%%%%%%%%%%%%%%%%%%%%%%%%%%%%%%%%%%%%%%%%%%%%%%%%

%%%%%%%%%%%%%%%%%%%%%%%%%%%%%%%%%%%%%%%%%%%%%%%%%%
\section{Introduction}
\label{sec:intro}
A spline of order $k$ is a piecewise  polynomial function
that obeys continuity conditions 
on its value and its first $k-2$ derivatives at the points, called {\it knots},
where the pieces join~\cite{deBoor}. Splines play an important role in nonparametric 
regression~\cite{wegman1983splines,wahba1990spline,hardle1990applied}, simply called 
curve fitting when the data
is one dimensional,
where the outcome is not assumed to have
a predetermined form of functional dependence on the predictor. 

It has long been recognized~\cite{wold1974spline,burchard1974splines,jupp1978approximation,luo1997hybrid} that 
 the quality of a spline fit 
 depends significantly on the locations of the knots defining the 
spline.
Determining the placement of knots that is best adapted to given data has proven to be 
a challenging non-linear and non-convex, not to mention high-dimensional, 
optimization problem that has resisted a satisfactory solution. 

A diverse set of methods have been proposed that either attempt this optimization problem head-on or solve an approximation to it in order to get a reasonable solution.
In the latter category, methods based on  
knot insertion and 
deletion~\cite{smith1982curve,lyche1988data,friedman1989flexible,friedman1991,stone1997polynomial} have been
studied extensively. In these methods, one starts with a fixed set of sites for knots
and performs a step-wise addition or removal of knots at these sites. The 
best number of 
knots  is determined by  
a model selection criterion such as Generalized Cross Validation (GCV)~\cite{luo1997hybrid,golub1979generalized}. Step-wise 
change in knot placement 
is not an efficient
exploration of the continuous space of possible knot 
positions and the end result, while computationally inexpensive to obtain and tractable
to mathematical analysis,
is not necessarily the best possible~\cite{zhou2001spatially}. Another
approach explored in the literature is the two-stage framework in which 
the first stage identifies a subset of active or dominant knots and the 
second stage merges them in a data dependent way to obtain a reduced set of knots~\cite{park2007b,kang2015knot,luo2019knot}. These methods have shown good performance for low noise applications.
% To compensate, a larger number of knots is needed than a completely free placement may actually require. 

In attempts at solving the optimization challenge directly, general purpose stochastic optimization algorithms (metaheuristics) such as Genetic Algorithm (GA)~\cite{geneticalgo}, Artificial Immune System (AIS)~\cite{ulker2009automatic} or
those based on Markov Chain Monte Carlo (MCMC)~\cite{green1995reversible}, have 
been studied~\cite{pittman2002adaptive,dimatteo2001bayesian,YOSHIMOTO2003751,miyata2003adaptive}. These methods have proven quite successful
in solving many challenging high-dimensional optimization problems in other fields and 
it is only natural to employ them for the problem of free knot placement.
However, GA and AIS are more suited to discrete optimization problems
rather than the  
inherently continuous one in knot optimization, and MCMC is 
computationally expensive.  Thus, there is plenty of scope for using 
other metaheuristics to find better solutions.

It was shown in~\cite{galvez2011efficient}, and independently in~\cite{mohanty2012particle}, that Particle Swarm Optimization (PSO)~\cite{PSO},
a relatively recent 
entrant to the field of nature inspired metaheuristics such as GA, 
is a promising method for the free knot placement problem. PSO is governed by a much 
smaller set of parameters than GA or MCMC and most of these do not 
appear to require much tuning from one problem to another. In fact, 
as discussed later in the paper, essentially two parameters 
are all that need to be explored to find a robust operating point 
for PSO.

An advantage of free knot placement is that 
a subset of knots can move close enough  
to be considered as a single knot with a higher
multiplicity. 
A knot with multiplicity $> 1$  can be used to construct splines that can  fit curves with discontinuities. 
Thus, allowing knots 
to move and merge opens up the possibility of modeling even 
non-smooth curves. 
That PSO can handle regression models requiring 
knot merging was demonstrated in~\cite{galvez2011efficient} albeit for 
examples with very low noise levels.  

% A question that has not been addressed adequately so far, either with addition-deletion schemes or with metaheuristics, is that of performance of adaptive spline fitting
% on data with a high noise level. There is a growing need for extending non-parametric regression to highly noisy 
% data in many scientific fields. This is especially acute in the
% case of gravitational wave (GW) searches that have recently achieved
% success~\cite{abbott2019gwtc}  (and the 2017 Nobel prize in Physics). The most important target for GW searches, where the data is a noisy time series, are transient chirps: signals whose instantaneous frequency, $\omega(t)$, and amplitude, $a(t)$, evolve adiabatically on the timescale of the instantaneous period, $1/\omega(t)$. For 
% GW sources having a sufficiently accurate theoretical model for 
% predicting $\omega(t)$ and $a(t)$, parametric regression methods such as matched filtering~\cite{Helstrom} can be used 
% and the associated 
% data analysis problem is well under control (except for computational issues).
% However, if either $\omega(t)$ or $a(t)$, or both, are not known a priori, non-parametric regression methods must be used and the high level of noise in GW data
% makes this
% an especially challenging task.
% In this context, the use of a spline-based non-parametric regression method (SEECR) was 
% shown in~\cite{SEECR-PhysRevD.96.102008} to be much more effective than searches based on time-frequency analysis. In SEECR, both $\omega(t)$ and $a(t)$ are modeled by splines, and knot optimization using PSO is used for fitting $\omega(t)$.

It was found in~\cite{SEECR-PhysRevD.96.102008}, and later in a simplified model problem~\cite{mohanty2018swarm}, that the  advantage 
engendered by free knot placement turns
into a liability as the level of noise increases: knots 
can form spurious clusters to 
fit outliers arising from noise, producing spikes in the resulting estimate and making 
it worse than useless. This problem was found to be mitigated~\cite{mohanty2018swarm}
by introducing a suitable regulator~\cite{ruppert2003semiparametric}. Regularization
has also been used in combination with knot addition~\cite{luo1997hybrid} but its role there -- suppression of
numerical 
instability arising from a  large 
numbers of knots -- is very different. 

% While the coalescence of knots in the free knot placement approach can allow modeling of 
% discontinuous curves, it should be avoided if there is prior knowledge that the true 
% curve is free of discontinuities. A novel penalty, based on the distance between knots, was 
% introduced in~\cite{lindstrom1999penalized} for suppressing coalescence of knots.   

% The progress described above in adaptive spline fitting using free knot placement  has happened over decades and in somewhat isolated steps that were often
% limited by the available computing power. 
The progress on free knot placement described above  has happened over decades and in somewhat isolated steps that were often
limited by the available computing power.
However, 
the tremendous growth in
computing power and the development of more powerful metaheuristics has finally 
brought us to the doorstep of a satisfactory resolution of this problem, at least
for one-dimensional regression.

In this paper, we  combine PSO based knot placement with regularization 
into a single algorithm  for adaptive spline 
fitting.
The algorithm, called Swarm Heuristics based Adaptive and Penalized
 Estimation of Splines (\fitalgoname), has  the flexibility to
 fit non-smooth functions as well as smooth ones without any change in 
 algorithm settings. It uses model selection to determine the best number 
 of knots, and reduces estimation
 bias arising from the regularization
 using a least squares 
 derived rescaling.  Some of the elements of \fitalgoname outlined above were explored in~\cite{mohanty2018swarm} in the context of a single
 example with a simple and smooth function. However, the crucial feature of allowing knots to merge  was missing there along with the step of bias reduction.
 (The bias reduction step does not seem to have been used elsewhere to the best of 
 our knowledge.)
 
 Various design choices involved in \fitalgoname
 are identified clearly and their effects are examined using large-scale 
 simulations and a diverse set of benchmark functions. Most importantly, 
\fitalgoname is applied to data with a much higher noise level than has traditionally been 
considered in the field of adaptive spline fitting and found to have promising performance. This sets the stage for further development of the adaptive spline
methodology for new application domains.

The rest of the paper is organized as follows.
 Sec.~\ref{sec:splinefit} provides a brief review of pertinent topics in spline 
 fitting. The PSO metaheuristic and the particular variant used in this paper are 
 reviewed in Sec.~\ref{sec:pso}. Details of \fitalgoname are described in Sec.~\ref{sec:fitalgorithm} along with the principal design choices. The 
 setup used for our simulations is described in Sec.~\ref{sec:simulation_setup}.
 Computational 
 aspects of \fitalgoname
 are addressed in Sec.~\ref{sec:compcosts}. This is followed by the presentation of results in Sec.~\ref{sec:results}.  Our conclusions are summarized in Sec.~\ref{sec:conclusions}. 
%%%%%%%%%%%%%%%%%%%%%%%%
\section{Fitting splines to noisy data}
\label{sec:splinefit}
In this paper, we consider the one-dimensional 
regression problem 
\begin{eqnarray}
y_i & = & f(x_i) + \epsilon_i\;,
\label{eq:regressionModel}
\end{eqnarray}
$i = 0,1,\ldots,N-1$, $x_0 = 0$, $x_{N-1} = 1$, $x_{i+1}>x_i$, with $f(x)$ 
unknown and
$\epsilon_i$ drawn independently from 
$N(0,1)$. The task is to find an estimate $\widehat{f}(x)$, given $\{y_i\}$,
of $f(x)$. 

% The ordinary least-squares solution for $f(x)$, $f(x_i) = y_i$, is not unique
% unless the 
% problem is regularized by restricting the space of 
% possible solutions. 
% One reasonable approach is to enforce ``smoothness"
% on the estimate $\widehat{f}$ of
% $f$. This can be done by choosing $\widehat{f}$ to be the minimizer 
% of the least-squares function but with
% a penalty on its average curvature: higher average curvature implying sharper turning points in
%$\widehat{f}$ and more ``roughness". Thus,
To obtain a non-trivial solution, the estimation problem must be regularized by 
restricting $\widehat{f}(x)$ to some specified class of functions.
One reasonable approach is to require that this be the class of ``smooth" functions, and
obtain the estimate as the solution of the variational problem,
\begin{eqnarray}
\widehat{f} & = & \argmin_f \left[\sum_{i=0}^{N-1}\left(y_i - f(x_i)\right)^2 + \lambda \int_0^1 dx \left(f^{\prime\prime}(x)\right)^2\right]\;.
\label{eq:smoothing_spline}
\end{eqnarray}
It can be shown that the solution   belongs to the space of 
cubic splines defined by $\{x_i\}$ as the set of knots. 
Consequently, $\widehat{f}$ is known as the {\it smoothing spline} estimate~\cite{wahba1990spline,reinsch1967smoothing}.
In Eq.~\ref{eq:smoothing_spline}, the first term on the right measures the fidelity of the model to the observations and the second term penalizes the ``roughness", measured by the average 
squared curvature, of the model.  The trade-off between these competing requirements is controlled by 
$\lambda\geq 0$, called the regulator gain or smoothing parameter. 

The best choice for $\lambda$ is the principle issue in practical applications of smoothing spline.
The use of GCV to adaptively determine the  value of $\lambda$ was introduced in~\cite{craven1978smoothing} and is used, for example, in the implementation of smoothing spline in the R~\cite{R-software} \texttt{stats} package. A
 scalar $\lambda$, adaptively selected or otherwise, is not well suited to handle a function with a heterogeneous roughness distribution across its domain. 
 The use of a spatially adaptive gain function, $\lambda(x)$, has been investigated in different forms~\cite{wahba_donoho2002wavelet,storlie2010locally,liu2010data,wang2013smoothing} to 
 address this issue.

% It is empirically observed that the large number of knots -- all the values 
% in $\{x_i\}$ -- in spline smoothing results in a
% high degree of freedom and overfitting.
A different regularization approach is to eschew an explicit penalty term and 
regularize the fitting problem by restricting the 
number of knots to be $\ll N$. This leads to the {\em regression spline}~\cite{wold1974spline} estimate
in which $\widehat{f}(x)$ is represented as a linear combination of a finite set of basis functions -- the so-called B-spline functions~\cite{deBoor,curry1947spline} being a popular choice --
that span the space of splines associated with the chosen knot sequence and polynomial order.
Different methods for adaptive selection of the number of knots, which is the main free parameter in regression spline, have been compared in~\cite{wand2000comparison}.
 The asymptotic properties of smoothing and regression spline 
 estimates have been analyzed theoretically in~\cite{claeskens2009asymptotic}.
 
% Restricting the number of knots to be $\ll N$ is one way to control
%  overfitting and this leads to the {\em regression spline}~\cite{wold1974spline} estimate. 
%  Analysis
%  of the asymptotic properties of smoothing and regression spline 
%  estimates also lends support to this modification~\cite{claeskens2009asymptotic}.

Smoothing and regression splines are hybridized in the {\em penalized spline}~\cite{ruppert2003semiparametric,eilers1996flexible,eilers2015twenty} approach: the deviation of
the spline model from the data is measured by the least squares function as in the first term of Eq.~\ref{eq:smoothing_spline} but the penalty becomes a quadratic form in the coefficients of the 
spline in the chosen basis set.  As in the case of smoothing spline, adaptive selection of the scalar regulator gain can be performed using GCV~\cite{ruppert2003semiparametric} and locally adaptive gain coefficients have been proposed in~\cite{ruppert2000theory,krivobokova2008fast,scheipl2009locally,yang2017adaptive}. The performance of alternatives to GCV for selection of a scalar regulator gain have been investigated and compared in~\cite{krivobokova2013smoothing}. 

While penalized spline is less sensitive to the number of knots, it is still
a free parameter of the algorithm that must be specified. Joint 
adaptive selection of the number of knots and regulator gain 
has been investigated in~\cite{luo1997hybrid,ruppert2002selecting} using GCV. Other model selection
methods can also be used for adaptive determination of the number of knots (see Sec.~\ref{sec:modelSelect}).

% Sec.~\ref{sec:intro}
% reviewed the other major determinant, besides regulator gain and number of knots, of the performance of regression and penalized spline: the placement of the knots. The remainder of this section describes regression and penalized splines along with their 
% generalization using free knots.

%%%%%%%%%%%%%%%%%
\subsection{B-spline functions}
\label{sec:B-splines}
Given a set of $M$ knots $\overline{b} = (b_0,b_1,\ldots,b_{M-1})$, $b_i\in [0,1]$, $b_{i+1}>b_i$, and given order $k$ of the 
spline polynomials, the set of splines 
that interpolates $\{(y_i,b_i)\}$, $y_i \in \mathbb{R}$, 
forms a linear vector space of dimensionality $M+k-2$. 
A convenient choice for a basis of
this vector space is the set of B-spline functions~\cite{curry1947spline}.

In this paper, we need B-spline functions for
the more general case of  a knot sequence $\overline{\tau}= (\tau_0,\tau_1,\ldots,\tau_{P-1})$, $\tau_{i+1}\geq \tau_i$ with $P>M$ knots, in which a
knot can appear more than once. The number of repetitions of any knot cannot be greater than $k$. Also, $\tau_j=b_0$ for $0\leq j\leq k-1$, and $\tau_j = b_{M-1}$ for $P-k\leq j\leq P-1$. The span of 
B-spline functions defined over a 
knot sequence with repetitions can contain functions that have jump  
discontinuities in their values or in their derivatives.
(The dimensionality of the span is $P-k$.)

\begin{figure}
     \centering
     \includegraphics[scale=0.45]{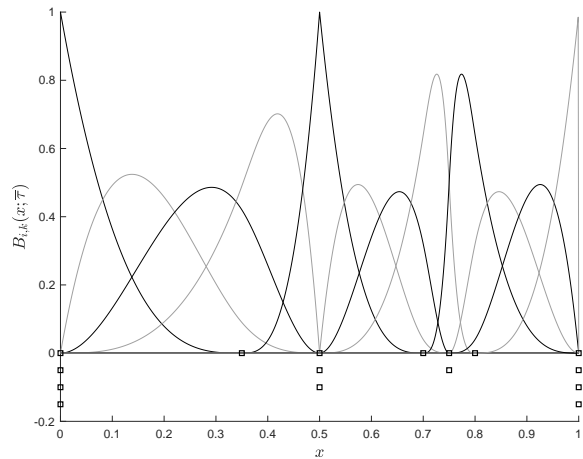}
     \caption{Cubic B-spline functions $\{B_{i,4}(x;\overline{\tau})\}$, $i = 0,1,\ldots,11$, for an 
     arbitrary choice of $16$ knots ($\overline{\tau}$) marked by squares.
     For visual clarity, alternate 
     B-spline functions are shown in black and gray. 
     Knots with multiplicity $> 1$ result in B-splines that
     are discontinuous in value or derivatives.}
     \label{fig:bsplinefig}
 \end{figure}

The Cox-de~Boor  
 recursion relations~\cite{DEBOOR197250} given below 
 provide an efficient way to compute the set of B-spline functions,
 $\{B_{i,k}(x;\overline{\tau})\}$, for any given order.
 The recursions start with B-splines of order 1, which are piecewise constant functions
\begin{eqnarray}
  B_{j,1}(x;\overline{\tau}) & = & \left\{\begin{array}{cc}
      1,  &  \tau_j \leq x < \tau_{j+1} \\
      0 & {\rm else}
  \end{array}\right.\;.
  \label{charfunc}
\end{eqnarray}
For $2\leq k^\prime \leq k$,
\begin{eqnarray}
B_{j,k^\prime}(x) & = & \omega_{j,k^\prime}(x) B_{j,k^\prime-1}(x) + \gamma_{j+1,k^\prime}(x) B_{j+1,k^\prime-1}(x)\;,\label{deboor_original}\\
\omega_{j,k^\prime}(x) & = & \left\{
\begin{array}{cc}
\frac{x-\tau_j}{\tau_{j+k^\prime-1}-\tau_j}\;, & \tau_{j+k^\prime-1}\neq \tau_j\\
0\;, & \tau_{j+k^\prime-1}= \tau_j
\end{array}
\right.\;,\\
\gamma_{j,k^\prime}(x) & = & \left\{
\begin{array}{cc}
1 - \omega_{j,k^\prime}(x)\;, & \tau_{j+k^\prime-1}\neq \tau_j\\
0\;, & \tau_{j+k^\prime-1}= \tau_j
\end{array}
\right.\;.
\end{eqnarray}
 In the recursion above,  $0\leq j \leq P - k^\prime -1$. Fig.~\ref{fig:bsplinefig} provides an illustration of B-spline functions. 
 
The regression spline method is elegantly formulated in 
terms of B-spline functions. The estimate is assumed to 
belong to the parametrized family of linearly combined B-spline functions,
\begin{eqnarray}
f(x;\overline{\alpha},\overline{\tau}) &=& \sum_{j = 0}^{P-k-1}\alpha_j B_{j,k}(x_i;\overline{\tau})\;,
\label{eq:fittingFunction}
\end{eqnarray}
where $\overline{\alpha} = (\alpha_0,\alpha_1,\ldots,\alpha_{P-k-1})$. The 
least-squares estimate is given by $\widehat{f}(x) = f(x;\widehat{\alpha},\widehat{\tau})$, where $\widehat{\alpha}$ and $\widehat{\tau}$
minimize
\begin{eqnarray}
L(\overline{\alpha},\overline{\tau}) & = &
 \sum_{i=0}^{N-1} 
\left(y_i -f(x_i;\overline{\alpha},\overline{\tau})
\right)^2\;.
\end{eqnarray}

%%%%%%%%%%%%%%%%%%%%%%%%%
\subsection{Regression and penalized spline with free knot placement}
\label{sec:pspline_freeknts}
The penalized spline estimate is found by minimizing
\begin{eqnarray}
% L_\lambda(\overline{\alpha},\overline{\tau}) & = & L(\overline{\alpha},\overline{\tau}) + \lambda \sum_{j=0}^{P-k-1} \alpha_j^2\;,
L_\lambda(\overline{\alpha},\overline{\tau}) & = & L(\overline{\alpha},\overline{\tau}) + \lambda R(\overline{\alpha})\;,
\label{eq:penalizedSpline}
\end{eqnarray}
over the spline coefficients (c.f. Eq.~\ref{eq:fittingFunction}), where $R(\overline{\alpha})$ is the penalty,  while keeping the 
number of knots and knot locations fixed. In this paper, we choose
\begin{eqnarray}
R(\overline{\alpha}) & = & \sum_{j=0}^{P-k-1} \alpha_j^2\;,
\label{eq:SHPS_penalty}
\end{eqnarray}
for reasons explained below.
% thereby enforcing more smoothness on the solution. 
% When $L_\lambda(\overline{\alpha},\overline{\tau})$ is minimized keeping $\overline{\tau}$
% fixed, the method is called {\em penalized spline}. 

 Formally, the penalty function can be derived by substituting Eq.~\ref{eq:fittingFunction} in the roughness penalty. This would lead to a quadratic form similar to the penalty in Eq.~\ref{eq:SHPS_penalty} but with a kernel matrix that is not the identity matrix~\cite{ramsay_silverman_functional}. The elements of this matrix would be Euclidean inner products of B-spline derivatives. However, using such a penalty 
 adds a substantial computational burden in free knot placement because it has to be recomputed every time the knot placement changes. Computational aspects of this problem are discussed in~\cite{eilers1996flexible}, where a simplified form of the roughness penalty is used that is based on the differences of coefficients of adjacent B-splines. This is a good approximation for the case considered in~\cite{eilers1996flexible} of a large number of fixed knots and closely spaced B-splines, but not necessarily for free knots that may be small in number and widely spread out. Another
 perhaps more important consideration is that repeated knots in free knot placement result in
 B-splines with discontinuous derivatives. This makes the  
 kernel matrix particularly challenging for numerical evaluation and increases code complexity. 
In this paper, we avoid the above issues by using the simple form of the penalty function in Eq.~\ref{eq:SHPS_penalty} and leave the investigation of more appropriate forms to future work. We note that the exploration of innovative penalty functions is an active topic of research (e.g.,~\cite{lindstrom1999penalized,eilers2015twenty,goepp2018spline}). 

While the reduction of the number of knots in regression spline coupled with 
the explicit regularization of penalized spline reduces overfitting, the fit is 
now sensitized to where the knots are placed.
Thus, the complete method involves the minimization of $L_\lambda(\overline{\alpha},\overline{\tau})$ (c.f., Eq.~\ref{eq:penalizedSpline})
over
both $\overline{\alpha}$ and $\overline{\tau}$. (The method of 
regression spline with 
knot optimization and explicit regularization will be referred to as {\em adaptive spline} in the following.)

Minimization of $L_\lambda$ over $\overline{\alpha}$ and $\overline{\tau}$ can be nested as follows.
\begin{eqnarray}
\min_{\overline{\tau},\overline{\alpha}} L_\lambda(\overline{\alpha},\overline{\tau}) & = & \min_{\overline{\tau}}\left(
\min_{\overline{\alpha}}
L_\lambda(\overline{\alpha},\overline{\tau})
\right)\;.
\end{eqnarray}
The solution, $\widehat{\alpha}(\overline{\tau})$, of the inner minimization is expressed 
 in terms of the $(P-k)$-by-$N$ matrix ${\bf B}(\overline{\tau})$, with
\begin{eqnarray}
B_{m,n}(\overline{\tau}) & = & B_{m,k}(x_n;\overline{\tau})\;,
\end{eqnarray}
as
\begin{eqnarray}
\widehat{\alpha}(\overline{\tau}) & = & \overline{y}{\bf B}^T
{\bf G}^{-1}\;,\label{eq:alphahat}\\
{\bf G} & = & {\bf B}{\bf B}^T + \lambda {\bf I}\;,
\end{eqnarray}
where ${\bf I}$ is the $(P-k)$-by-$(P-k)$ identity matrix. The outer minimization
over $\overline{\tau}$ of
\begin{eqnarray}
F_\lambda(\overline{\tau}) & = & L_\lambda(\widehat{\alpha}(\overline{\tau}),\overline{\tau})\;,
\label{eq:regsplpsplfitfunc}
\end{eqnarray}
needs to be performed numerically. 

Due to the fact that freely moveable knots can coincide, and that this produces discontinuities in B-spline functions as outlined earlier,  curve fitting by adaptive spline can accommodate a 
broader class of functions -- smooth with localized discontinuities --
than smoothing or penalized spline. 

The main bottleneck in implementing the 
adaptive spline method is the global minimization of $F_\lambda(\overline{\tau})$ since it is 
 a high-dimensional non-convex function having multiple 
local minima.  Trapping by local minima renders greedy methods
ineffective and high dimensionality makes a brute force search for the
global minimum 
computationally infeasible. 
This is where PSO enters the picture and, as shown later,
offers a way forward.

%%%%%%%%%%%%%%%%%%%%
\subsection{Model selection}
\label{sec:modelSelect}
In addition to the parameters $\overline{\alpha}$ and $\overline{\tau}$, adaptive spline has two hyper-parameters, namely the 
regulator gain $\lambda$ and the number of interior knots $P-2(k-1)$, that affect the outcome of fitting. Model selection methods can be employed
to fix these hyper-parameters based on the data.

In this paper, 
we restrict ourselves to the adaptive selection of only the number of knots. 
This is done by 
minimizing the Akaike Information Criterion (AIC)~\cite{akaike1998information}: For a regression model
with $K$ parameters $\overline{\theta}=(\theta_0,\theta_1,\ldots,\theta_{K-1})$, 
\begin{eqnarray}
{\rm AIC} & = & 2K - 2\max_{\overline{\theta}}\ln\left(\Lambda(\overline{\theta})\right)\;, 
\end{eqnarray}
where $\Lambda(\overline{\theta})$ is the likelihood function. 
The specific 
expression for AIC used in \fitalgoname is provided in Sec.~\ref{sec:fitalgorithm}. 

%%%%%%%%%%%%%%%%%%%%%%%%%%%%%%%%%%%%%%%
\section{Particle swarm optimization}
\label{sec:pso}
Under the PSO metaheuristic, the
function to be optimized (called the {\em fitness function}) is 
 sampled at a fixed number of locations (called {\em particles}). 
 The set of particles is called a {\em swarm}.
 The particles 
move in the search space following stochastic iterative 
rules called {\em dynamical equations}. The dynamical equations implement
two essential features called {\em cognitive} and {\em social} forces.  They
serve to retain ``memories" of the best locations found by the particle and
the swarm (or a subset thereof) respectively.

Since its introduction by Kennedy and Eberhart~\cite{PSO}, 
 the PSO metaheuristic has expanded to include a large diversity of algorithms~\cite{engelbrecht2005fundamentals}.
In this paper, we consider the variant called
 local-best (or {\em lbest}) PSO~\cite{lbestTopology}. 
 We begin with the notation~\cite{normandin2018particle}
  for describing lbest PSO.
\begin{itemize}
    \item $F(x)$: the scalar  fitness function to be minimized, with 
     $x = (x^1, x^2, \ldots, x^d)\in \mathbb{R}^d$. 
    In our case, $x$ is $\overline{\tau}$, $F$ is  $F_\lambda(\overline{\tau})$ (c.f., Eq.~\ref{eq:regsplpsplfitfunc}),  and $d=P-2(k-1)$.
    \item $\mathcal{S}\subset \mathbb{R}^d$: the search space
     defined by the hypercube $a^j\leq x^j \leq b^j$, $ = 1, 2, \ldots, d$
     in which the global minimum of the fitness function must be found.
    \item $N_p$: the number of particles in the swarm.
    \item $x_i[k]\in \mathbb{R}^d$: the position of the $i^{\rm th}$ particle 
    at the $k^{\rm th}$ iteration.
    \item $v_i[k]\in \mathbb{R}^d$: a vector called the {\em velocity} of the $i^{\rm th}$ particle that is used for updating the position of a particle.
    \item $p_i[k]\in \mathbb{R}^d$: the best location found by the $i^{\rm th}$ particle over all iterations up to and including the $k^{\rm th}$. $p_i[k]$ is called the {\em personal best} position
    of the $i^{\rm th}$ particle.
    \begin{eqnarray}
        F(p_i[k]) & = &\min_{1\leq j\leq k} F(x_i[j])\;.
    \end{eqnarray}
    \item $n_i[k]$: a set of particles, called the {\em neighborhood} of particle $i$, $n_i[k]\subseteq 
    \{1,2,\ldots,N_p\}\setminus \{i\}$. There are many possibilities, called {\em topologies}, for
       the choice of $n_i[k]$. In the simplest, called the global best topology, 
       every particle is the neighbor of every other particle: $n_i[k]= 
    \{1,2,\ldots,N_p\}\setminus \{i\}$. The topology used for lbest PSO in this paper is described later.
    \item $l_i[k]\in \mathbb{R}^d$: the best location among the particles in $n_i[k]$  over all iterations up to and including the $k^{\rm th}$.
                    $l_i[k]$ is called the {\em local best}  for the $i^{\rm th}$ particle.
    \begin{eqnarray}
        F(l_i[k]) & = & \min_{j \in \{i\} \cup n_i[k]} F(p_j[k])\;.
    \end{eqnarray}
    \item $p_g[k]\in \mathbb{R}^d$: The best location among all the particles in the swarm,  $p_g[k]$ is called the {\em global best}.
        \begin{eqnarray}
        F(p_g[k]) & = & \min_{1\leq i \leq N_p} F(p_i[k])\;.
        \end{eqnarray}
\end{itemize}

The dynamical equations for {\em lbest} PSO are as follows.
\begin{eqnarray}
v_i[k+1] & = & w[k] v_i[k] + c_1  (p_i[k] - x_i[k]) {\bf r}_1 +\nonumber\\
            &&   c_2 (l_i[k] - x_i[k]) {\bf r}_2 \;,
            \label{velocityEqn}\\
x_i[k+1] & = & x_i[k] + z_i[k+1]\;,\label{eq:positionEqn}\\
z^j_i[k] & = & \left\{
\begin{array}{cc}
  v_i^j[k]\,,  &  -v_{\rm max}^j \leq v_i^j[k] \leq v_{\rm max}^j\\
   -v_{\rm max}^j\,,  & 
   v_i^j[k] < -v_{\rm max}^j\\
   v_{\rm max}^j\,, & v_i^j[k] > v_{\rm max}^j
\end{array}
\right.\;.
\label{eq:velocityClamping}
\end{eqnarray}
Here, 
$w[k]$ is a deterministic function known as the 
inertia weight, $c_1$ and $c_2$
are constants, and ${\bf r}_i$ is a diagonal matrix with 
iid random variables having a
uniform distribution 
over $[0,1]$. Limiting the velocity as shown in Eq.~\ref{eq:velocityClamping} is
called {\it velocity clamping}.

The iterations are initialized at $k=1$ 
by independently drawing (i)  $x_i^j[1]$ 
from a uniform distribution over $[a^j,b^j]$, and (ii) 
$v_i^j[1]$ from a uniform distribution over $[a^j-x_i^j[1],b^j-x_i^j[1]]$.
For termination of the iterations, we use the simplest condition: terminate when a prescribed number ${\niter}$ of iterations are completed. 
The solutions found by PSO for the minimizer and the minimum value of
the fitness 
are $p_g[\niter]$ and $F(p_g[\niter])$ respectively. Other, more sophisticated,
termination conditions are available~\cite{engelbrecht2005fundamentals}, but 
the simplest one has served well across a variety of regression problems in our 
experience.

The second and third terms on the RHS of Eq.~\ref{velocityEqn} are 
the cognitive and social forces respectively. On average they attract  
a particle towards its personal and local bests, promoting 
the exploitation of an already good solution to find better ones nearby.
The term containing the inertia weight, on the other hand,  
promotes motion along the same direction and allows a particle to 
resist the cognitive and social forces. Taken together, the terms control the
exploratory and exploitative behaviour of the algorithm.
We allow the inertia weight $w[k]$ to decrease linearly with $k$
from an initial value $w_{\rm max}$ to a final value $w_{\rm min}$
in order to transition PSO from an 
 initial exploratory to a final exploitative phase.
 
For the topology, we use the {\em ring topology} with $2$
       neighbors in which
\begin{eqnarray}
       n_i[k] = \left\{\begin{array}{cc}\{i-1,i+1\}\;,& i \notin \{1,N_p\}\\
                                        \{N_p,i+1\}\;, & i = 1\\
                                        \{i-1,1\}\;, & i = N_p
       \end{array}\right.\;.
\end{eqnarray}
The local best, $l_i[k]$, in the $k^{\rm th}$ iteration is updated after evaluating the fitnesses of all the particles. The velocity and position updates
given by Eq.~\ref{velocityEqn} and Eq.~\ref{eq:positionEqn} respectively form the last set of operations in the $k^{\rm th}$ iteration. 

 To handle particles that exit the
search space, we use the ``let them fly" 
boundary condition under which a particle outside the search space is 
assigned a fitness value of $\infty$. Since both $p_i[k]$ and 
$l_i[k]$ are always within the search space, such a particle is
eventually pulled back into the search space by the 
cognitive and social forces. 

%%%%%%%%%%%%%%%%%%%%%%%%%%%%%%%%%%
\subsection{PSO tuning}
\label{tuning}
Stochastic global optimizers, including PSO, that terminate in a finite
number of iterations do not
satisfy the conditions laid out in~\cite{solis1981minimization} for  
convergence to the global
optimum. Only the probability of convergence can be improved by 
tuning the parameters of the algorithm for a given optimization problem.

In this sense, most of the parameters involved in PSO are found to have 
fairly robust values when tested across an extensive suite of 
benchmark fitness functions~\cite{bratton2007defining}.
Based on widely prevalent values in the literature, these are: 
 $N_p = 40$, 
$c_1 = c_2 =  2.0$, $w_{\rm max} = 0.9$,
$w_{\rm min} = 0.4$, and $v_{\rm max}^j = 0.5[b^j - a^j]$. 

Typically, this leaves the
maximum number of iterations, $\niter$, as the principal
parameter that needs to be tuned. However, 
for a given $\niter$, the probability of 
convergence can be increased by the simple strategy of running 
multiple, independently initialized 
runs of PSO on the same fitness function and choosing the 
best fitness value found across the runs. The probability of missing the global optimum decreases 
exponentially as $(1-P_{\rm conv})^{\nruns}$, where $P_{\rm conv}$ is
the probability of successful convergence in any one run and $\nruns$ is
the number of independent runs. 

Besides $\niter$, therefore,  $\nruns$ is the remaining parameter that should be 
tuned. If the independent runs can be parallelized, $\nruns$ 
is essentially fixed by the available number of parallel workers although  
this should not be stretched to the extreme. If too high a value of $\nruns$ is needed in an application (say $\nruns \geq 8$), it is usually an indicator that
$P_{\rm conv}$
should be increased by tuning the other PSO parameters or by exploring a different
PSO variant. In this paper,  we follow the simpler way of tuning $\nruns$ 
by setting it to $\nruns = 4$,  the typical number of processing cores available in a high-end desktop.

%%%%%%%%%%%%%%%%%%%%%%%%%%%%%%%%%%%%%%%%%%%%%
\section{\fitalgoname algorithm}
\label{sec:fitalgorithm}
The \fitalgoname algorithm is summarized in the 
pseudo-code given in Fig.~\ref{fig:pseudocode}. The user specified parameters
of the algorithm are (i) the number, $\nruns$, of PSO to use per data realization; 
(ii) the number of iterations, $\niter$, to termination of PSO; (iii) the set of 
models, $\mathbb{N}_{\rm knots}$,
 over which AIC based model selection (see below) is used; (iv) the regulator gain $\lambda$. Following the standard initialization condition for PSO (c.f., Sec.~\ref{sec:pso}), the initial knots for each run of PSO  are drawn independently from a uniform distribution over $[0,1]$.  
\begin{figure}
\begin{algorithmic}
\State {\bf Input:}
\State $\bullet$ $\overline{y}\gets$ Data
\State $\bullet$ $\nruns\gets$ Number of PSO runs
\State $\bullet$ $\niter\gets$ Maximum number of iterations
\State $\bullet$ $\mathbb{N}_{\rm knots}\gets \{M_1,M_2,\ldots,M_{\rm max}\}$; Number of knots (not counting repetitions) 
\State $\bullet$ $\lambda \gets$ Regulator gain
\State {\bf Execute:}
\For{$M\in \mathbb{N}_{\rm knots}$}\Comment Loop over models
    \For{$r\in\{1,2,\ldots,\nruns\}$} \Comment (Parallel) loop over PSO runs
        \State $\widehat{\tau}(r)\gets $ $\argmin_{\overline{\tau}}F_\lambda(\overline{\tau})$ using PSO \Comment Best location
        \State $\widehat{\alpha}(r)\gets$ B-spline coefficients corresponding to $\widehat{\tau}(r)$
        \State $F(M,r)\gets$ $F_\lambda(\widehat{\tau}(r))$  \Comment Best fitness value
    \EndFor
    \State $r_M\gets \argmin_{r} F(M,r)$ \Comment Best PSO run
    \State ${\rm AIC}(M)\gets {\rm AIC}$ for $F(M,r_M)$ (c.f., Eq.~\ref{eq:aic4shapes}) 
    \State $\widehat{f}(M)\gets $ Estimated function corresponding to $\widehat{\tau}(r_M)$ and $\widehat{\alpha}(r_M)$
\EndFor
\State $M_{\rm best}\gets \argmin_M {\rm AIC}(M)$ \Comment Model with lowest AIC
\State $\widehat{f}\gets$ $\widehat{f}(M_{\rm best})$
\State $\widehat{f}\gets$ Bias corrected $\widehat{f}$ (c.f., Sec.~\ref{sec:bias}) 
\State {\bf Output:}
\State $\bullet$ $M_{\rm best}$\Comment Best model
\State $\bullet$ Estimated, bias-corrected $\widehat{f}$ \Comment Estimated function from best model
\State $\bullet$ $F(M_{\rm best},r_{M_{\rm best}})$ \Comment Fitness of best model
\end{algorithmic}
\caption{Pseudo-code for the \fitalgoname algorithm. All quantities 
with parenthesized integer arguments stand for arrays, with the argument as the
array index. }
\label{fig:pseudocode}
\end{figure}

A model in \fitalgoname is specified by
the number of non-repeating knots.  For each model $M\in \mathbb{N}_{\rm knots}$,  $F(M,r_M)$ denotes the fitness value, where $1\leq r_M\leq \nruns$ is  the best PSO run.
The AIC value for the model is  given by 
\begin{eqnarray}
{\rm AIC} & = & 4 M + F(M,r_M)\;,
\label{eq:aic4shapes}
\end{eqnarray}
which follows from the number of optimized parameters being $2M$ (accounting
for both knots and B-spline coefficients) and the log-likelihood   being proportional
to the least squares function for the noise model used here.
(Additive constants that 
do not affect the minimization of AIC have been dropped.)

The algorithm acts on given data $\overline{y}$ to produce (i) the best fit model $M_{\rm best}\in \mathbb{N}_{\rm knots}$; (ii) the fitness value associated
 with the best fit model; (iii) the estimated function $\widehat{f}$ from the best fit model. The generation of $\widehat{f}$ includes a bias correction step 
 described next.
 
%%%%%%%%%%%%%%%%%%%%%%%%
\subsection{Bias correction}
\label{sec:bias}
The use of a non-zero regulator gain  
leads to shrinkage in the estimated B-spline coefficients. As a result, 
the corresponding 
estimate, $\widehat{f}$, has a systematic point-wise bias towards zero. 
A bias correction transformation is applied to $\widehat{f}$ as follows.

First, the unit norm estimated function $\widehat{u}$
is obtained, 
\begin{eqnarray}
\widehat{u} & = & \frac{\widehat{f}}{\|\widehat{f}\|}\;,
\end{eqnarray}
where $\|\widehat{f}\| = [\sum \widehat{f}_i^2]^{1/2}$ is the $L_2$ norm.

Next, a scaling factor $A$ is estimated as
\begin{eqnarray}
A & = & \argmin_a \sum_{i=0}^{N-1}\left(y_i - a \widehat{u} \right)^2\;.
\end{eqnarray}
The final estimate is given by $\widehat{f} = A\widehat{u}$.

As discussed earlier in Sec.~\ref{sec:pspline_freeknts} (c.f. Eq.~\ref{eq:penalizedSpline} and Eq.~\ref{eq:SHPS_penalty}), the 
penalty used in this paper is one among 
several alternatives available in the literature. For some
forms of the penalty, there need not be any shrinkage in the B-spline coefficients and the bias correction step above would be unnecessary.
%%%%%%%%%%%%%%%%%%%%%%
\subsection{Knot merging and dispersion}
\label{sec:knotmerging}
In both of the mappings described in Sec.~\ref{sec:knotMap},
it is possible to get knot sequences
in which a subset $(\tau_i,\tau_{i+1},\ldots,\tau_{i+m-1})$ of $1< m \leq M-2$ 
of interior knots falls within an interval $(x_j,x_{j+1})$ between two consecutive
predictor values.  
There are two possible options to handle such a situation. 
\begin{itemize}
    \item{Heal}: Overcrowded knots are dispersed such that there is only one knot  between any two consecutive predictor values. This can be done iteratively by moving a knot to the right or left depending on the difference in distance to the corresponding neighbors. 
    \item{Merge}: All the knots in an overcrowded set are made equal to the rightmost knot  $\tau_{i+m-1}$ until its multiplicity saturates at $k$. The remaining knots,  $\tau_i$ to $\tau_{i+m-1-k}$, are equalized to the remaining rightmost knot $\tau_{i+m-1-k}$ until its multiplicity staturates to $k$, and so on. (Replacing rightmost by leftmost when merging is an equally valid alternative.) Finally, if more than one set of merged knots remain within
    an interval $(x_j,x_{j+1})$, they are dispersed by healing. 
\end{itemize} 
If only healing is used, \fitalgoname
cannot fit curves that have jump discontinuities in value or derivatives. Therefore, 
if it is known that the unknown curve in the data is free of jump discontinuities, 
 healing acts as an implicit regularization to enforce this condition. Conversely, merging should be used when jump discontinuities cannot be 
 discounted.

It is important to note that in both healing and merging, the number of 
knots stays fixed at $M+2(k-1)$ where $M\in \mathbb{N}_{\rm knots}$.

%%%%%%%%%%%%%%%%%%%%
\subsection{Mapping particle location to knots}
\label{sec:knotMap}
For a given model $M\in \mathbb{N}_{\rm knots}$, the search space for PSO
is $M$ dimensional.
 Every particle location, $\overline{z} = (z_0,z_1,\ldots,z_{M-1})$, 
 in this space has to be mapped to 
an $M+2(k-1)$ element knot sequence $\overline{\tau}$ before evaluating 
its fitness
$F_\lambda(\overline{\tau})$. 

We consider two alternatives for the map from $\overline{z}$ to $\overline{\tau}$. 
\begin{itemize}
    \item{Plain}:  $\overline{z}$ is sorted in ascending order. After sorting, $k-1$ copies of $z_0$ and $z_{M-1}$ are prepended and appended respectively to $\overline{z}$. These are the repeated end knots as described in Sec.~\ref{sec:B-splines}.  
    \item{Centered-monotonic}: In this scheme~\cite{calvin-siuro}, the search space is the unit hypercube: $z_i \in [0,1]$, $\forall i$. First, an initial set of $M$ knots is obtained from
    \begin{eqnarray}
    z_0 & = & \tau_0\;,\\ 
    z_{1\leq i \leq M-2} & = & \frac{\tau_i - \tau_{i-1}}{\tau_{i+1}-\tau_{i-1}}\;,\\
    z_{M-1} & = & \frac{\tau_{M-1}-\tau_0}{1-\tau_0}\;.
    \end{eqnarray}
    This is followed by prepending and appending $k-1$ copies of $\tau_0$ and $\tau_{M-1}$ respectively to the initial knot sequence.
\end{itemize}
In the plain map,
any permutation of $\overline{z}$ maps into the same knot sequence
due to sorting.
This creates  degeneracy in $F_\lambda$, which
may be expected to make the task of global minimization harder for PSO. The centered-monotonic map 
is designed to overcome this problem: by construction, it assigns a unique
 $\overline{\tau}$ to a given $\overline{z}$. Moreover, $\overline{\tau}$ is 
 always a monotonic sequence, removing the need for a sorting operation. 
 This map also has the nice normalization
that the center
of the search space at $z_i = 0.5$, $1\leq i\leq M-2$, corresponds to uniform 
spacing of the interior knots. 

It should be noted here that the above two maps are not the only possible ones. 
The importance of the ``lethargy theorem" (degeneracy of the fitness function) and 
using a good parametrization for the knots in regression spline
was pointed out by Jupp~\cite{jupp1978approximation} back in $1978$. A logarithmic
map for knots was proposed in~\cite{jupp1978approximation} that, while not implemented 
in this paper, should be examined in future work.
%%%%%%%%%%%%%%%%%%%%%%%%
\subsection{Optimization of end knots}
\label{sec:endknotsfree}
When fitting curves to noisy one-dimensional 
data in a signal processing context, a common 
situation is that the signal is transient and localized well away from the end points $x_0$ and $x_{N-1}$
of the predictor. However, the location of the signal in the data -- its time of 
arrival in other words -- may be unknown. 
In such a case, it makes sense to keep the end knots free and
subject to optimization. 

On the other hand, if it is known that the curve occupies the entire
predictor range, it is best to fix the end
knots by keeping $z_0$ and $z_{M-1}$ fixed. (This reduces the dimensionality of the
search space for PSO by $2$.)

%%%%%%%%%%%%%%%%%%%%%%%
\subsection{Retention of end B-splines}
\label{sec:endBspline}
The same  signal processing scenario considered above
suggests that, for signals that decay smoothly to
zero at their start and end, it is best to drop the end B-spline functions because
they have a jump discontinuity in value (c.f., Fig~\ref{fig:bsplinefig}). 
In the contrary case, the end B-splines may be retained so that the estimated 
signal can 
start or end at non-zero values.

%%%%%%%%%%%%%%%%%%%%%%%%%%%%%%%%%%%%%%%%%%%%%%%%%%
\section{Simulation study setup}
\label{sec:simulation_setup}
We examine the performance of \fitalgoname
on simulated data with a wide range of benchmark functions. In this section, we
present these functions, the simulation
protocol used, the metrics for quantifying
performance, and a scheme 
for labeling test cases that is used in Sec.~\ref{sec:results}.  (In the following, 
the terms ``benchmark function" and ``benchmark signal" are used interchangeably.)

%%%%%%%%%%%%%%%%%%%%%%%%%
\subsection{Benchmark functions}
\begin{figure}
    \centering
    \includegraphics[scale=0.45]{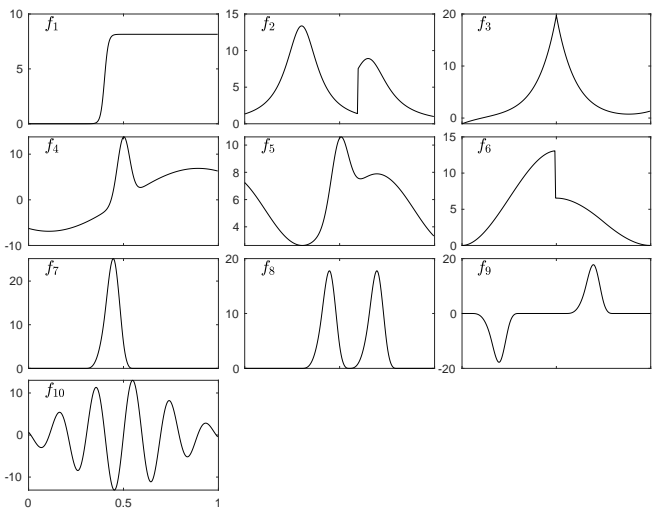}
    \caption{Benchmark functions normalized to have $\snr=100$. 
    The function name is 
    indicated in the upper left corner of each panel. The abscissa in
    each panel is identical to the one showing $f_{10}$. }
    \label{fig:allBFsigPlots}
\end{figure}

The benchmark functions used in this study are listed in Table~\ref{tab:benchmarkFunctions} and 
plotted in Fig.~\ref{fig:allBFsigPlots}.  
%%%%%%%%%%%%%%%% ARXIV PREPRINT %%%%%%%%%%%%%%%%%%%%%
 \squeezetable{
%%%%%%%%%%%%%%%%%%%%%%%%%%%%%%%%%%%%%%%%%%%%%%%%%%%%%
\begin{table}
    \caption{The benchmark functions used in this paper. 
    The sources from which the functions have been obtained are: $f_1$ to $f_3$ \cite{YOSHIMOTO2003751}; $f_4$~\cite{dimatteo2001bayesian}; $f_5$~\cite{denison1998automatic,li2006bayesian}; $f_6$~\cite{lee2002algorithms}; $f_7$~\cite{mohanty2018swarm}. Functions  $f_8$ to $f_{10}$ are introduced here.
        \label{tab:benchmarkFunctions}
    }
%%%%%%%%%%%%%%%% ARXIV PREPRINT %%%%%%%%%%%%%%%%%%%%%
 \begin{ruledtabular}
%%%%%%%%%%%%%%%%%%%%%%%%%%%%%%%%%%%%%%%%%%%%%%%%%%%%%
    \centering
    \renewcommand{\arraystretch}{1.5}
    %%%%%%%%%%%%%%%% ARXIV PREPRINT %%%%%%%%%%%%%%%%%%%%%
     \begin{tabular}{|l|c|}
    %%%%%%%%%%%%%%%%%%%%%%%%%%%%%%%%%%%%%%%%%%%%%%%%%%%%%
    %%%%%%%%%%%%%%%%% COMPTSTATS (SPRINGER) %%%%%%%%%%%%%%%%%
    %  \begin{tabular}{lc}
    % \hline\noalign{\smallskip}
    %%%%%%%%%%%%%%%%%%%%%%%%%%%%%%%%%%%%%%%%%%%%%%%%%%%%%%%%%
    {\bf Expression} & {\bf Domain} \\
    %%%%%%%%%%%%%%%%% COMPTSTATS (SPRINGER) %%%%%%%%%%%%%%%%%
    % \noalign{\smallskip}\hline\noalign{\smallskip}
    %%%%%%%%%%%%%%%%%%%%%%%%%%%%%%%%%%%%%%%%%%%%%%%%%%%%%%%%%
    %%%%%%%%%%%%%%%% ARXIV PREPRINT %%%%%%%%%%%%%%%%%%%%%
     \hline
    %%%%%%%%%%%%%%%%%%%%%%%%%%%%%%%%%%%%%%%%%%%%%%%%%%%%%
    $ f_1(x) = 90 (1+e^{-100(x-0.4)})^{-1}$     & $x \in [0, 1]$ \\ \hline
    $
        f_2(x)=\begin{cases}
          (0.01+(x-0.3)^2)^{-1} \\
          (0.015 + (x-0.65)^2)^{-1} \\
     \end{cases}
    $ & $\begin{aligned}0\leq x < 0.6 \\ 0.6 \leq x \leq 1\end{aligned}$ \\ \hline
    
    $f_3(x) = 100 e^{-|10x-5|} + (10x-5)^5/500$ & $x \in [0, 1]$ \\  \hline
    $f_4(x) = \sin(x) + 2e^{-30x^2}$ & $x \in [-2, 2]$  \\  \hline
    $f_5(x) = \sin(2x) + 2e^{-16x^2} + 2$ & $x \in [-2, 2]$  \\  \hline
    
    $f_6(x) = \begin{cases}
        4x^2(3-4x) \\
        \frac{4}{3}x(4x^2-10x+7)-\frac{3}{2} \\
        \frac{16}{3}x(x-1)^2
        \end{cases}$ & $\begin{aligned}
            0 \leq x < 0.5 \\
            0.5 \leq x < 0.75 \\
            0.75 \leq x \leq 1
        \end{aligned}$ \\ \hline
       $\begin{array}{rc}
    f_7(x) =  B_{3,4}(x;\overline{\tau})\; ;& \overline{\tau} = (\tau_0,\tau_1,\ldots,\tau_{11})\\
    \tau_i & =\left\{ \begin{array}{cc}
                      0.3 \;,& 0\leq i \leq 2\\
                      0.55 \;,& 8\leq i \leq 10
    \end{array}    \right.\\
    (\tau_3,\ldots,\tau_7) & = (0.3,0.4,.45,0.5,0.55)
    \end{array}$ & $x\in [0,1]$ \\  \hline
    
    $f_8(x) = B_{3,4}(x;\overline{\tau})+B_{3,4}(x-0.125;\overline{\tau})$
    & $x\in [0,1]$ \\ \hline
    
    $f_9(x) = B_{3,4}(x-0.25,\overline{\tau})+B_{3,4}(x-0.125;\overline{\tau})$
    & $x\in [0,1]$  \\ \hline
    
    $f_{10}(x) = e^{-\frac{(x-0.5)^2}{0.125}}\sin\left(10.24\pi (x-0.5)\right)$
    & $x\in [0,1]$ 
%%%%%%%%%%%%%%%%% COMPTSTATS (SPRINGER) %%%%%%%%%%%%%%%%%
    %\\ \noalign{\smallskip}\hline
%%%%%%%%%%%%%%%%%%%%%%%%%%%%%%%%%%%%%%%%%%%%%%%%%%%%%%%%%%
    \end{tabular}
%%%%%%%%%%%%%%%% ARXIV PREPRINT %%%%%%%%%%%%%%%%%%%%%
     \end{ruledtabular}
%%%%%%%%%%%%%%%%%%%%%%%%%%%%%%%%%%%%%%%%%%%%%%%%%%%%%
\end{table}
%%%%%%%%%%%%%%%% ARXIV PREPRINT %%%%%%%%%%%%%%%%%%%%%
 }
%%%%%%%%%%%%%%%%%%%%%%%%%%%%%%%%%%%%%%%%%%%%%%%%%%%%%

Function $f_1$ has a sharp change but is differentiable everywhere. Functions
$f_2$ and $f_6$ have jump discontinuities, and $f_3$ has a jump discontinuity in
its slope. Functions $f_4$ and $f_5$ are smooth but  
sharply peaked. 
Functions $f_7$ to $f_{10}$ all decay to zero at both ends and
serve to model smooth but transient signals; $f_7$ to
$f_9$ are designed to require progressively higher number of knots for fitting;
$f_{10}$ is an oscillatory signal that is typical for signal processing applications and expected to require the highest number of knots.
In addition, $f_7$ and $f_8$ test the ability of \fitalgoname to localize
time of arrival.

%%%%%%%%%%%%%%%%%%%%%%%%%
\subsection{Data simulation}
Following the regression model in Eq.~\ref{eq:regressionModel}, a simulated 
data realization 
consists of pseudorandom iid noise drawn from $N(0,1)$ added to a given
benchmark function that is sampled uniformly at 256 points in $[0,1]$. 

We consider the performance of \fitalgoname 
across a range of signal to noise ratio ($\snr$) defined as,
\begin{eqnarray}
\snr & = & \frac{\|f\|}{\sigma}\;,
\label{eq:snr}
\end{eqnarray}
where $f$ is a benchmark function and $\sigma$ is the standard deviation -- set to unity in this paper --
of the noise.
For
each combination of benchmark function and $\snr$,  \fitalgoname is applied
to $N_R=1000$ independent data realizations. This results in $1000$ corresponding 
estimated functions. Statistical summaries, such as the point-wise mean and standard deviation of the estimate, are computed from this set of estimated functions.

%%%%%%%%%%%%%%%%%%%%%%%%%
\subsection{Metrics}
The principal performance metric used in this paper is the sample root mean squared error (RMSE):
\begin{eqnarray}
{\rm RMSE} & = & \left[\frac{1}{N_R}\sum_{j=1}^{N_R}\| f - \widehat{f}_j\|^2\right]^{1/2}\;,
\end{eqnarray}
where $f$ is the true function in the data 
and $\widehat{f}_j$ its estimate from the $j^{\rm th}$ data realization. 
We use bootstrap with $10^4$ independently
drawn samples with replacement from the set $\{\| f - \widehat{f}_j\|^2\}$
to obtain the sampling error in RMSE.

A secondary metric that is useful is the sample mean of the number of knots 
 in the best fit model. To recall, this is the average of $M_{\rm best}\in\mathbb{N}_{\rm knots}$ over the $N_R$ data realizations,
 where $M_{\rm best}$ and $\mathbb{N}_{\rm knots}$ were
 defined in Sec.~\ref{sec:fitalgorithm}. The error in $M_{\rm best}$ is estimated by its sample standard deviation.

%%%%%%%%%%%%%%%%%%%%%%%%%
\subsection{Labeling scheme}
Several design choices in \fitalgoname were described in Sec.~\ref{sec:fitalgorithm}.  A useful bookkeeping  device
for keeping track of the many possible
 combinations of these choices is the labeling scheme  presented in Table~\ref{tab:labelScheme}.  
\begin{table}
    \centering
    \begin{tabular}{|c|c|c|}
    \hline
        PSO algorithm (Sec.~\ref{sec:pso})& $L$: lbest PSO &  $\ast$    \\ \hline
       \multirow{2}{*}{Knot Map (Sec.~\ref{sec:knotMap})}  
                               & $P$:     & $C$:\\
                              & Plain     & Centered-monotonic \\ \hline
        $\snr$ (Eq.~\ref{eq:snr})             & (Numerical) &  \\ \hline
        $\lambda$ (Eq.~\ref{eq:penalizedSpline}) & (Numerical) & \\ \hline
        $\niter$ (Number of PSO iterations) & (Numerical) & \\ \hline
        End knots  (Sec.~\ref{sec:endknotsfree}) & $F$: Fixed & $V$: Variable \\ \hline
        End B-splines (Sec.~\ref{sec:endBspline}) & $K$: Keep & $D$: Drop \\ \hline
        Knot merging (Sec.~\ref{sec:knotmerging}) & $M$: Merge & $H$: Heal\\
        \hline
    \end{tabular}
    \caption{Labeling scheme for a combination of design choices in \fitalgoname. The string labeling a combination is formed 
    by going down the rows of the
    table and (a) picking one letter from the last two columns of 
    each row, or (b) inserting 
    the value of a numerical quantity. 
    Numerical values in the key string are demarcated by underscores 
    on both sides. Thus, a key string looks like $Y_1Y_2\_X_3\_X_4\_X_5\_Y_6Y_7Y_8$ where 
    $Y_i$  and $X_i$ stand for letter and numerical entries respectively, and $i$ is the row number of the table starting from the top. We have left the possibility open for replacing lbest PSO with some other variant in the future. This is indicated by the `$\ast$' symbol in the top row.}
    \label{tab:labelScheme}
\end{table}

Following
this labeling
scheme, a string such as ${\rm LP}\_100\_0.1\_50\_{\rm FKM}$ refers to the combination:
lbest PSO; plain map from PSO search space to knots; $\snr=100$ for the true 
function in the data; regulator gain $\lambda = 0.1$; maximum number of
PSO iterations set to $50$; end knots fixed; end B-splines retained; merging
of knots allowed.

\section{Computational considerations}
\label{sec:compcosts}
The results in this paper were obtained with a code implemented entirely 
in {\sc Matlab}~\cite{matlab}. 
Some salient points about the code are described below.

The evaluation of B-splines uses the efficient algorithm given in~\cite{deBoor}. 
Since our current B-spline code
is not vectorized, it suffers a performance 
penalty in {\sc Matlab}. (We estimate that it is $\approx 50\%$ slower as a result.)
 Nonetheless, the code is reasonably fast: A single PSO run on a single data realization, for the more expensive case of $\snr=100$, takes about $11$~sec on an Intel Xeon (3.0 GHz)
class processor. It is important to note that
the run-time above is specific to the set, $\mathbb{N}_{\rm knots}$,
of models used.
In addition, 
due to the fact that the number of particles breaching the search space
boundary in a given PSO iteration is a random
variable and that the fitness of such a particle 
is not computed, the actual run times vary slightly
for different PSO runs and data realizations. 
 
The only parallelization used in the current code is over the independent PSO runs.
Profiling shows that $\approx 60\%$ of the run-time in a single PSO run 
is consumed by the evaluation of particle fitnesses,
out of which $\approx 45\%$ is 
spent in evaluating B-splines.  
Further substantial saving in run-time is, therefore,
possible if particle fitness evaluations are also parallelized.
This dual parallelization is currently not possible in the {\sc Matlab}
code but, given that we use $N_p=40$ particles, parallelizing all $N_p$
fitness evaluations can be expected to reduce the run-time by about an order of magnitude. However, 
realizing such a large number of parallel processes 
needs hardware acceleration using, for example, Graphics Processing Units. 

The operations count in the most time-consuming parts of the code (e.g., evaluating B-splines)
scales linearly with the length of the data. Hence, the projected ratios above
in run-time speed-up are not expected to change 
much with data length although the overall run-time will grow linearly. 

The pseudorandom number streams used for the simulated
noise realizations 
and in the PSO dynamical equations utilized built-in and well-tested 
default generators. 
The PSO runs were assigned independent pesudorandom streams that
were initialized, at the start of processing any data realization,
with the respective run number as the seed.
This (a) allows complete reproducibility of results for a given data realization,
and (b) does not breach the cycle lengths of the pseudorandom number generators
when processing a large number of data realizations. 

%%%%%%%%%%%%%%%%%%%%%%%%%%%%%%%%%%%%%%%%%%%%%%%%%%%
\section{Results}
\label{sec:results}
The presentation of results is organized as follows.
Sec.~\ref{sec:SNRHigh} shows single data realizations and estimates for 
a subset of the benchmark functions. Sec.~\ref{sec:regulatorEffect} analyzes the impact of the regulator gain $\lambda$
on estimation. 
Sec.~\ref{sec:SNR100} and Sec.~\ref{sec:SNR10} contain results for 
$\snr=100$ and $\snr=10$ respectively.
Sec.~\ref{sec:comp_bias_correct} shows the effect of the bias correction step described in Sec.~\ref{sec:bias} on the performance of \fitalgoname for both $\snr$ values.
In Sec.~\ref{sec:comp_susr_smsp}, we compare the performance of \fitalgoname with two well-established smoothing methods, namely, wavelet-based thresholding and shrinkage~\cite{donoho1995adapting}, and smoothing spline with adaptive selection of the regulator gain~\cite{craven1978smoothing}. The former follows an approach that does not
use splines at all, while the latter uses splines but avoids 
free knot placement. As such, they provide a good contrast to the approach followed in \fitalgoname. 

In all applications of \fitalgoname,   the set of models used was 
\begin{equation*}
    \mathbb{N}_{\rm knots}=\{5, 6, 7, 8, 9, 10, 12, 14, 16, 18\}\;.
\end{equation*} 
The spacing 
between the models is set 
 wider for higher knot numbers in order to reduce the computational 
 burden involved in processing a large number of data realizations.
 In an application involving just a few realizations, a denser spacing 
 may be used.

\begin{figure}
    \centering
    \includegraphics[scale=0.45]{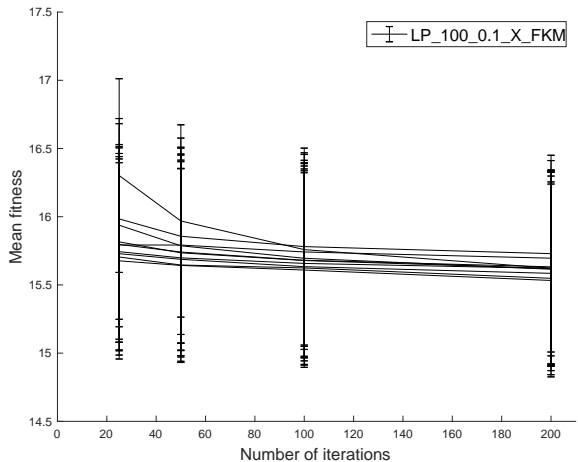}
    \caption{Performance of {\it lbest} PSO as a function of the 
    number of iterations, $\niter$, to termination. Each curve corresponds to 
    one of the benchmark functions at $\snr=100$ and shows 
    the mean fitness value as a 
    function of $\niter$. The mean fitness value is an average 
    over $N_R = 1000$ data realizations of the fitness value corresponding to the best model (i.e., $F(M_{\rm best},r_{M_{\rm best}})$ defined in Fig.~\ref{fig:pseudocode}).
    The error bars represent $\pm 1\sigma$
    deviations where $\sigma$ is the sample standard deviation.
    The other 
    algorithm settings used for this plot can be read off  from the key string 
    shown in the legend using Table~\ref{tab:labelScheme}.
    }
    \label{fig:pso_tuning}
\end{figure}

Fig.~\ref{fig:pso_tuning} shows the performance of lbest PSO 
across the set of benchmark functions as a function of the parameter
$\niter$. Given that the fitness values do not change in a statistically significant way
when going from $\niter = 100$ to $\niter=200$ in the SNR=100 case, we set it to the former as it
saves computational cost. A similar plot of fitness values
(not shown) for $\snr=10$ is used to set $N_{\rm iter} = 50$ for the $\snr=10$ case.
%PSO tuning using fitting function value

%%%%%%%%%%%%%%%%%%%%%%%%
\subsection{Sample estimates}
\label{sec:SNRHigh}
\begin{figure*}
%%%%%%%%%%%%%%%%%%%%%%%%%%%%%%%%%%%%%%%%%%%%%%%%%%%%%%%%%%%
    \centering

    \includegraphics[scale=0.6]{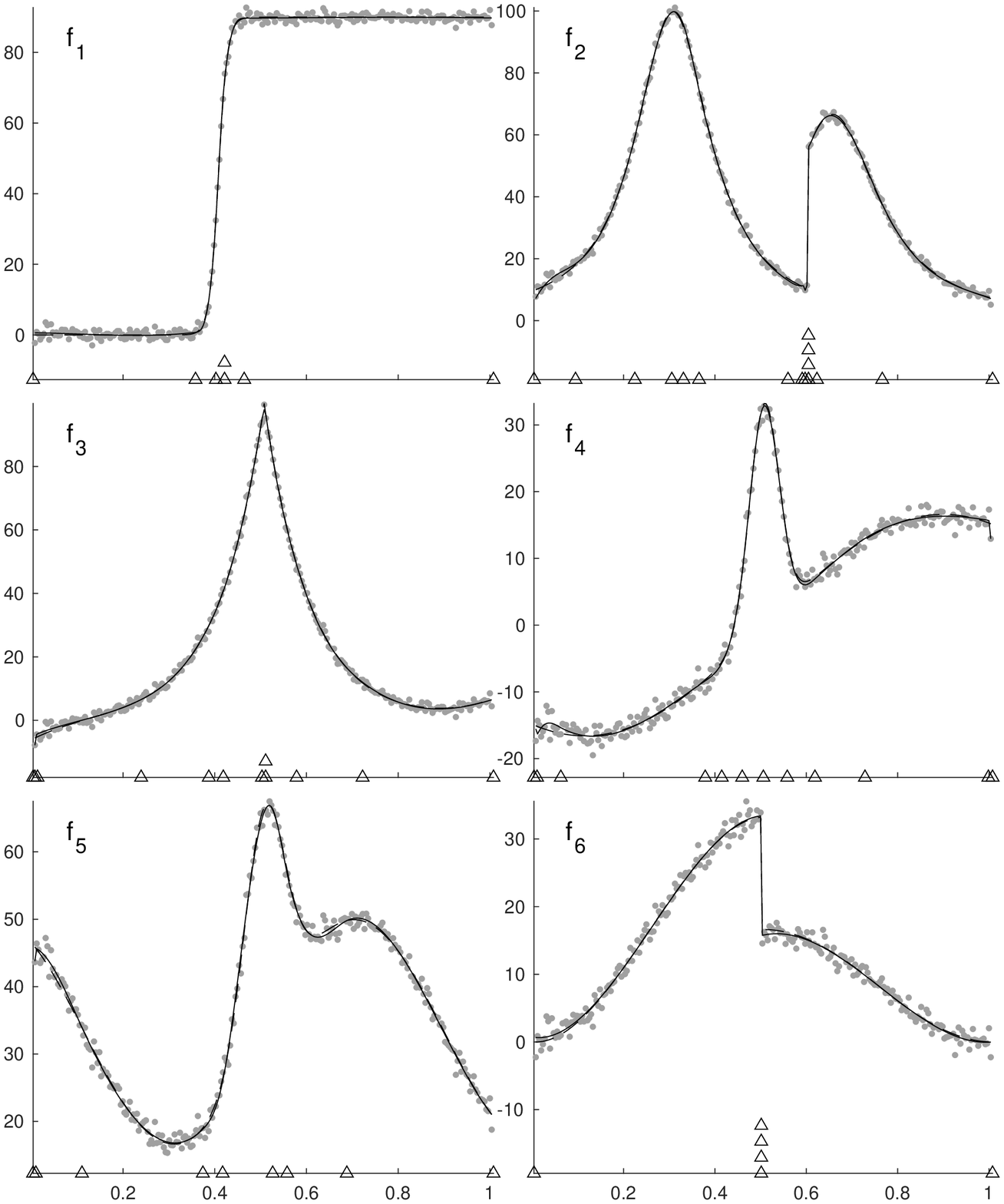}

     \caption{Sample estimated functions for the benchmark functions
    $f_1$ to $f_6$. 
    In each panel: the solid black curve is the estimated function; triangles show the locations of its knots (vertically stacked triangles denote repeated knots); the dashed black curve is the true function; gray dots represent the data realization.
    (In most cases, the solid and dashed curves are  visually indistinguishable.)
    The $\snr$s (rounded to integer values) of the functions in order from $f_1$ to $f_6$  are $1104$, $747$, $506$, $241$, $633$, and $254$, respectively. The algorithm settings were ${\rm LP}\_\snr\_0\_100\_{\rm FKM}$ (c.f., Table~\ref{tab:labelScheme}). Note that under these settings, the end B-splines are retained, which requires end knots to have the maximum allowed multiplicity. But this is a fixed multiplicity in each plot and not shown for clarity.
    }
    \label{fig:GI_Comp_estsigplot}
%%%%%%%%%%%%%%%%%COMPSTATS SPRINGER%%%%%%%%%%%%%%%%%%%%%%%%
% \end{figure}
%%%%%%%%%%%%%%%%%%%%%%%%%%%%%%%%%%%%%%%%%%%%%%%%%%%%%%%%%%%
%%%%%%%%%%%%%%%%%%ARXIV PREPRINT%%%%%%%%%%%%%%%%%%%%%%%%%%%
 \end{figure*}

In Fig.~\ref{fig:GI_Comp_estsigplot}, we show function estimates obtained 
with \fitalgoname for 
arbitrary single data realizations.  While not statistically rigorous, this allows
an initial assessment of performance  when the $\snr$ is sufficiently
high. Also shown with each estimate is the location of the knots found by \fitalgoname. 
%%%%%%%%%%%%%%%%%COMPSTATS SPRINGER%%%%%%%%%%%%%%%%%%%%%%%%
%\begin{figure}
%%%%%%%%%%%%%%%%%%%%%%%%%%%%%%%%%%%%%%%%%%%%%%%%%%%%%%%%%%%
%%%%%%%%%%%%%%%%%%ARXIV PREPRINT%%%%%%%%%%%%%%%%%%%%%%%%%%%
%%%%%%%%%%%%%%%%%%%%%%%%%%%%%%%%%%%%%%%%%%%%%%%%%%%%%%%%%%%

For ease of 
comparison, we have picked only  
the benchmark functions ($f_1$ to $f_6$) 
used in~\cite{galvez2011efficient}. The SNR of each function matches the value one would obtain using the noise standard deviation 
tabulated in~\cite{galvez2011efficient}.  Finally, the algorithm settings were brought as close as possible by (a) setting the regulator gain $\lambda = 0$, (b) using the plain map (c.f., Sec.~\ref{sec:knotMap}), (c) keeping the end knots fixed, and (d) allowing 
knots to merge. 
Differences remain in the PSO variant (and associated parameters) used and, 
possibly, the criterion used for merging knots.

We find that \fitalgoname has excellent 
performance at high $\snr$ values: without any change in settings, it can fit 
 benchmark functions ranging from spatially inhomogenous but smooth to ones that
 have discontinuities. For the latter, \fitalgoname allows knots to coalesce into repeated knots 
 in order to improve the fit at the location of the discontinuities. The sample estimates in Fig.~\ref{fig:GI_Comp_estsigplot} are visually indistinguishable from the ones given in~\cite{galvez2011efficient}. The same holds for the sample estimates given in~\cite{YOSHIMOTO2003751}, which uses benchmark functions $f_1$  to $f_3$ and SNRs similar to~\cite{galvez2011efficient}.

%%%%%%%%%%%%%%%%%%%%%%%
\subsection{Regulator gain}
\label{sec:regulatorEffect}
While the aim of restricting
the number of knots in regression spline is to promote a smoother estimate,
it is an implicit 
regularization that does not guarantee smoothness.
In the absence of an explicit regularization,  a fitting method based on free knot placement will exploit this loophole to form spurious
clusters of knots that fit 
outliers arising from noise and overfit the data. 
This issue becomes increasingly important 
as the level of noise in the data increases.

\begin{figure}
    \centering
    \includegraphics[scale=0.45]{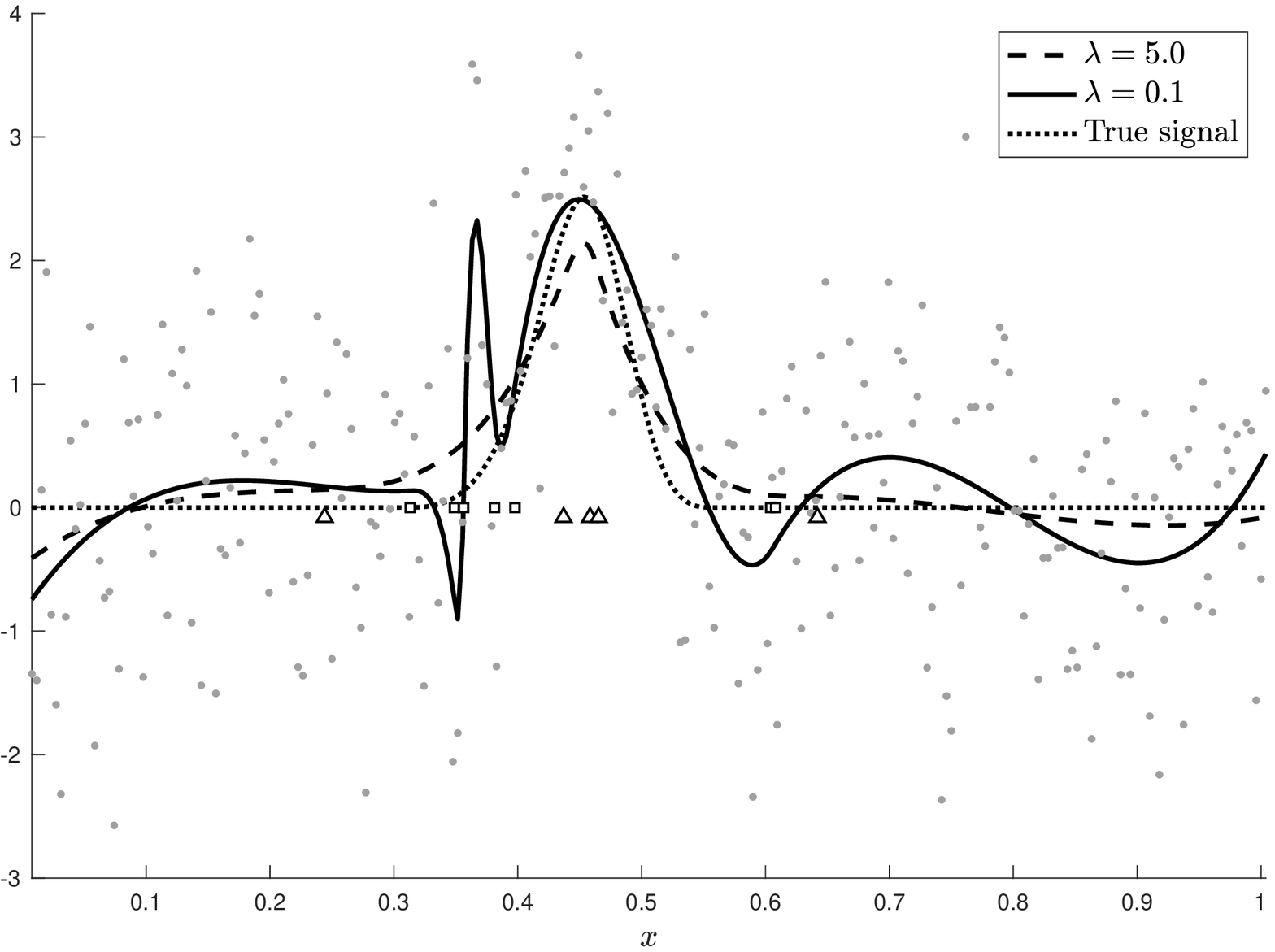}
    \caption{Effect of regulator gain, $\lambda$, on estimation. The solid and dashed curves are the estimates obtained with $\lambda = 0.1$ and $\lambda = 5.0$ respectively, where $\lambda$ is the regulator gain for the penalty term in Eq.~\ref{eq:penalizedSpline}. The true curve --  benchmark function $f_7$ with
    $\snr = 10$ -- is shown with a dotted line and the gray dots show the data realization. The interior knots in the best model 
     for $\lambda = 0.1$ and $\lambda = 5.0$ are shown as squares and triangles respectively. (Not shown here is an extra repeated knot for $\lambda = 0.1$.) Besides the difference in $\lambda$, the algorithm settings -- ${\rm LP}\_10\_\lambda\_50\_{\rm FKM}$ (see Table~\ref{tab:labelScheme}) -- were identical for the two estimated curves.
    }
    %LP_10_g_50_FKM, inFile_37
    \label{fig:regGainComp}
\end{figure}

Fig.~\ref{fig:regGainComp}
illustrates how adding the penalized spline regulator helps mitigate this problem
of knot clustering. Shown in the figure is one data realization and the 
corresponding estimates obtained with 
high and low values of the regulator gain $\lambda$.
For the latter, sharp spikes appear in the estimate 
where the function value is not high but the noise values are. The method 
tries to fit out these values by 
putting more knots in the model and clustering them to form the spikes.
Since knot clustering also needs large B-spline coefficients in order
to build a spike, a larger penalty
on the coefficients suppresses spurious spikes.

\begin{figure}
    \centering
    \includegraphics[scale=0.45]{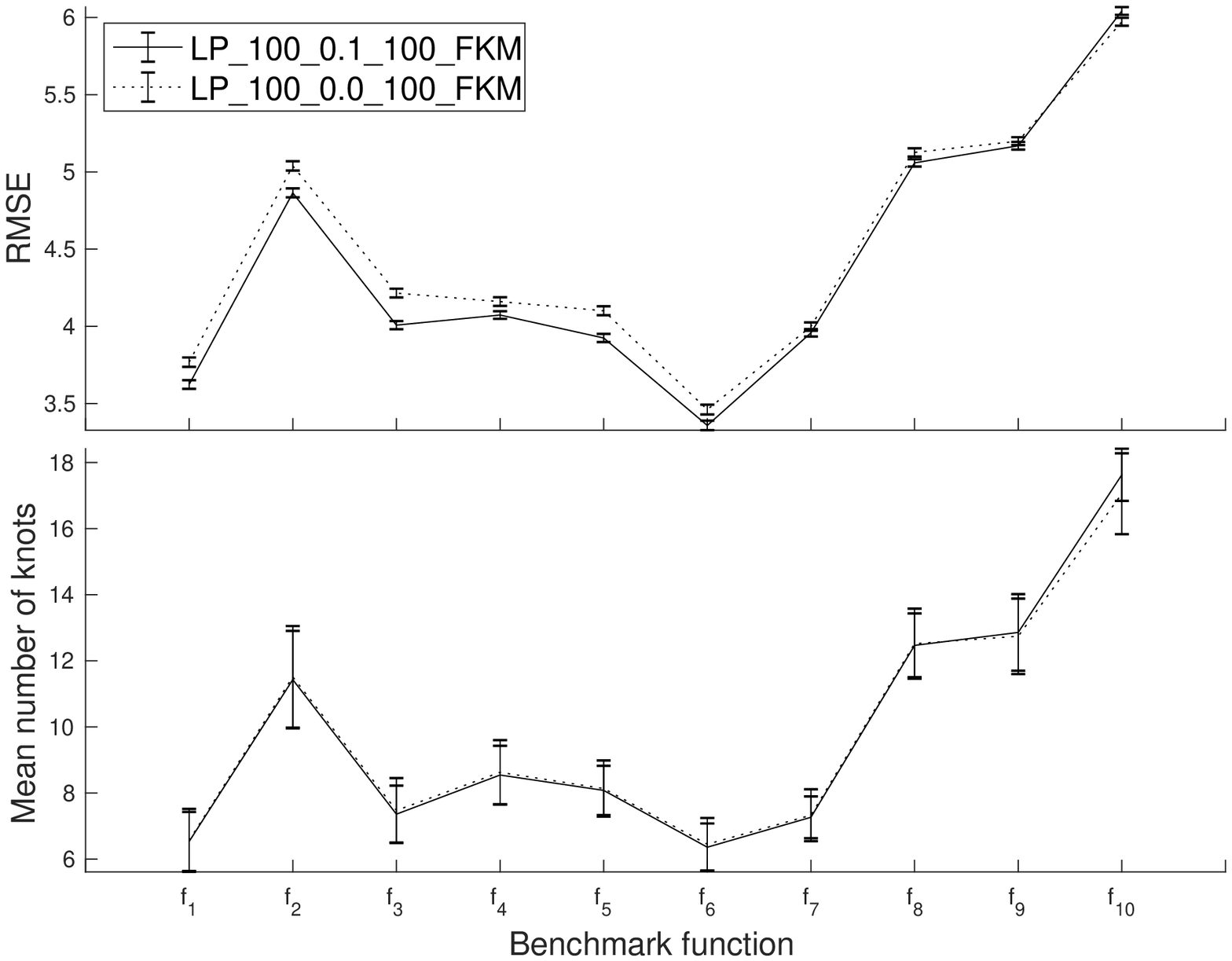}
    \caption{Effect of regulator gain on (top panel) the root mean squared error (RMSE), and  (bottom panel) the 
    mean number of knots in the best model for $\snr=100$ benchmark functions. In both panels, the solid and dotted curves correspond to $\lambda=0.1$ and $\lambda = 0.0$ respectively.   The other 
    algorithm settings used for this plot can be read off  from the key strings 
    shown in the legend using Table~\ref{tab:labelScheme}. The data points correspond to the benchmark functions shown on the abscissa. The error bars show $\pm 1 \sigma$
    deviations, where $\sigma$ is the estimated standard deviation.
    }
    \label{fig:SNR100_regGainComp}
\end{figure}

\begin{figure}
    \centering
    \includegraphics[scale=0.45]{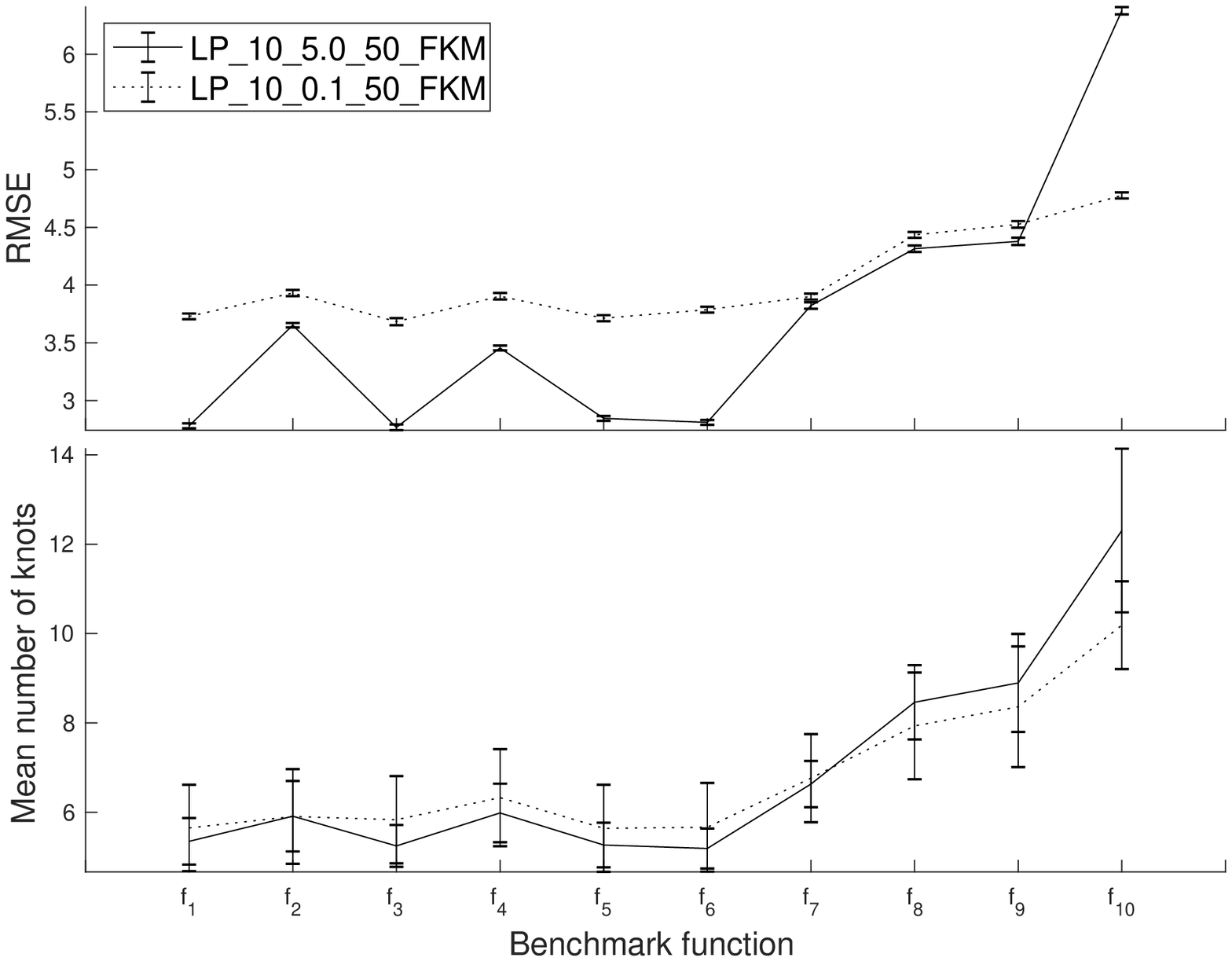}
    \caption{Effect of regulator gain on (top panel) the root mean squared error (RMSE), and  (bottom panel) the 
    mean number of knots in the best model for $\snr=10$ benchmark functions. In both panels, the solid and dotted
    curves correspond to $\lambda=5.0$ and $\lambda = 0.1$ respectively. The other
    algorithm settings used for this plot can be read off  from the key strings 
    shown in the legend using Table~\ref{tab:labelScheme}. The data points correspond to the benchmark functions shown on the abscissa. The error bars show $\pm 1 \sigma$
    deviations, where $\sigma$ is the estimated standard deviation.
    }
    \label{fig:SNR10_regGainComp}
\end{figure}

Fig.~\ref{fig:SNR100_regGainComp} and Fig.~\ref{fig:SNR10_regGainComp} 
present a statistically more rigorous study of the effect of $\lambda$
by examining the RMSE attained across the whole set of benchmark functions
at different $\snr$ values. In both figures, the RMSE is shown for identical
algorithm settings except for $\lambda$, and in both  we 
observe that increasing the regulator gain improves the RMSE. (The lone case where this is not true is addressed in more detail 
in Sec.~\ref{sec:SNR10}.) The
improvement becomes more pronounced as $\snr$ is lowered. 
(The effect of $\lambda$ on the number of knots in the best fit model at either
$\snr$ is within the sampling error of the simulation.)

 The higher values
 of the regulator gains in Fig.~\ref{fig:SNR100_regGainComp} and Fig.~\ref{fig:SNR10_regGainComp} -- $\lambda = 0.1$ and $\lambda = 5.0$ for $\snr=100$ and $\snr=10$  respectively -- 
 were chosen according to the $\snr$. These pairings were chosen empirically
 keeping in mind that there is an optimum regulator gain for a given noise level. Too high 
a gain becomes counterproductive as it simply shrinks the estimate towards
zero. Too low 
a value, as we have seen, brings forth the issue of knot clustering and spike formation.
Since the latter is a more serious issue for a higher noise level, the optimum regulator gain shifts towards a correspondingly higher value.

%%%%%%%%%%%%%%%%%%%%%%%%
\subsection{Results for $\snr=100$}
\label{sec:SNR100}
We have already selected some of the 
algorithm settings in the 
preceding sections, namely,  the number of 
iterations to use and the regulator gain
for a given $\snr$. Before proceeding further, we need to decide on the remaining
ones. 

For the $\snr=100$ case, it is clear that the end knots and end B-splines must
be retained because benchmark functions $f_1$ to $f_6$ do not all decay to zero and the noise level is not high enough to mask this behavior.
Similarly, knot merging is an obvious choice because discontinuities in some 
of the benchmark functions are obvious at this $\snr$ and they cannot 
be modeled under the alternative option of healing. 
The remaining choice is between
the two knot maps: plain or centered-monotonic. 

\begin{figure}
    \centering
    \includegraphics[scale=0.45]{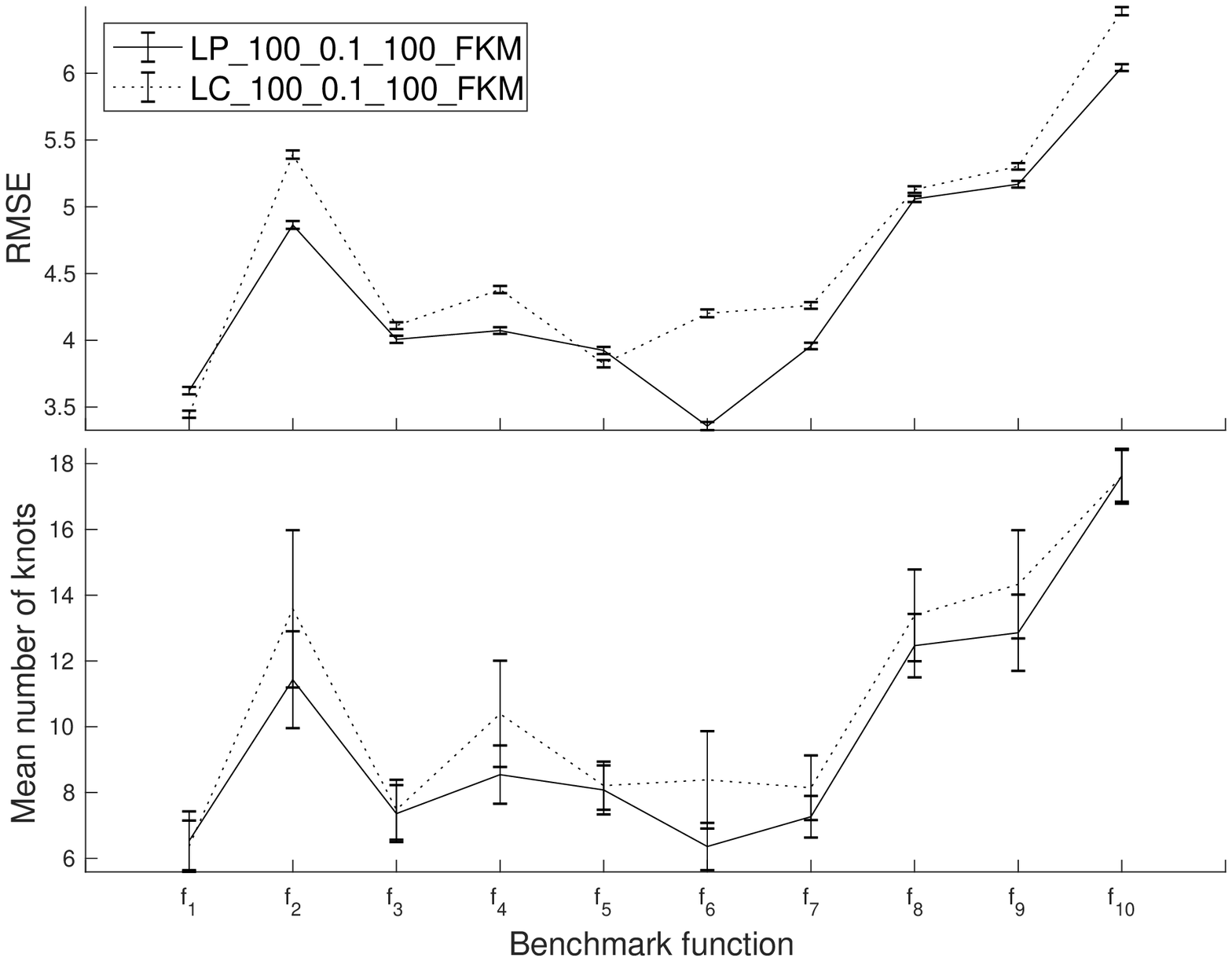}
    \caption{Effect of the map used for transforming PSO search space coordinates to knots on (top panel) the root mean squared error (RMSE), and (bottom panel) the 
    mean number of knots in the best model for $\snr=100$ benchmark functions.
    The solid and dotted curves correspond to the plain and centered-monotonic maps  respectively. The other
    algorithm settings used for this plot can be read off  from the key strings 
    shown in the legend using Table~\ref{tab:labelScheme}. The data points correspond to the benchmark functions shown on the abscissa. The error bars show $\pm 1 \sigma$
    deviations, where $\sigma$ is the estimated standard deviation.}
    \label{fig:SNR100_CompMaps}
\end{figure}

As shown in Fig.~\ref{fig:SNR100_CompMaps}, the RMSE is distinctly worsened by the centered-monotonic map across all the benchmark functions. This map also leads
to a higher number of knots in the best fit model although the difference
is not as significant statistically. Thus, the clear winner here is the map in 
which knots are merged.

\begin{figure}
    \centering
    \includegraphics[scale=0.45]{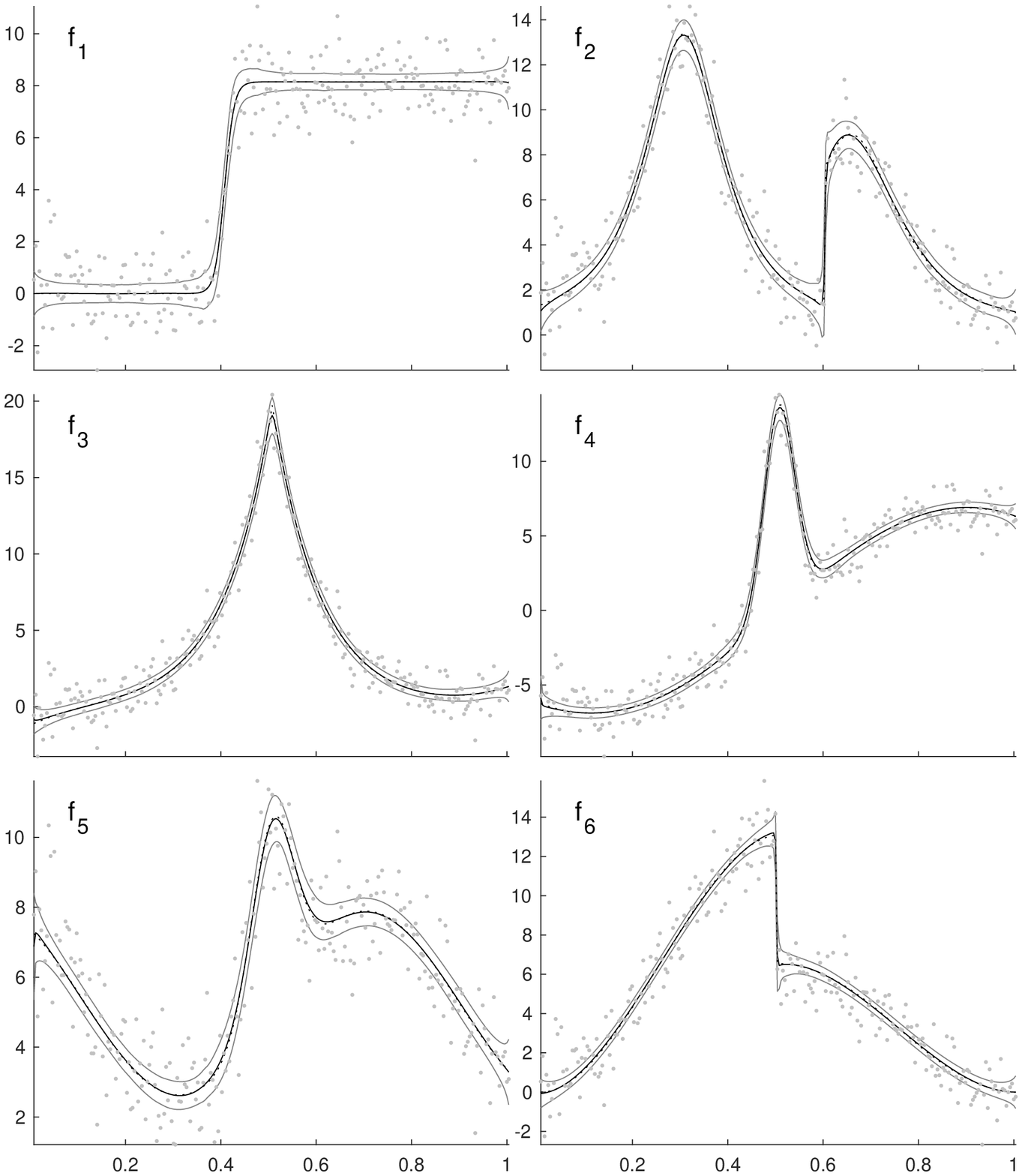}
    \caption{Mean estimated functions (black curve) for benchmark functions $f_1$ to $f_6$ at $\snr = 100$.  The true functions are shown as dotted curves
    but they are practically indistinguishable from the mean estimated functions.  The gray curves show $\pm 2\sigma$ 
    deviation from the mean function, where $\sigma$ is the estimated 
    standard deviation. The gray dots show an arbitrary 
    data realization for the purpose of visualizing the noise level.
     The abscissa has the same range for each panel. The algorithm settings used are given by
    the key string ${\rm LP}\_100\_0.1\_100\_{\rm FKM}$, which can be expanded using Table~\ref{tab:labelScheme}.
    }
    \label{fig:SNR100_MeanEstSig_1to6}
\end{figure}

\begin{figure}
    \centering
    \includegraphics[scale=0.45]{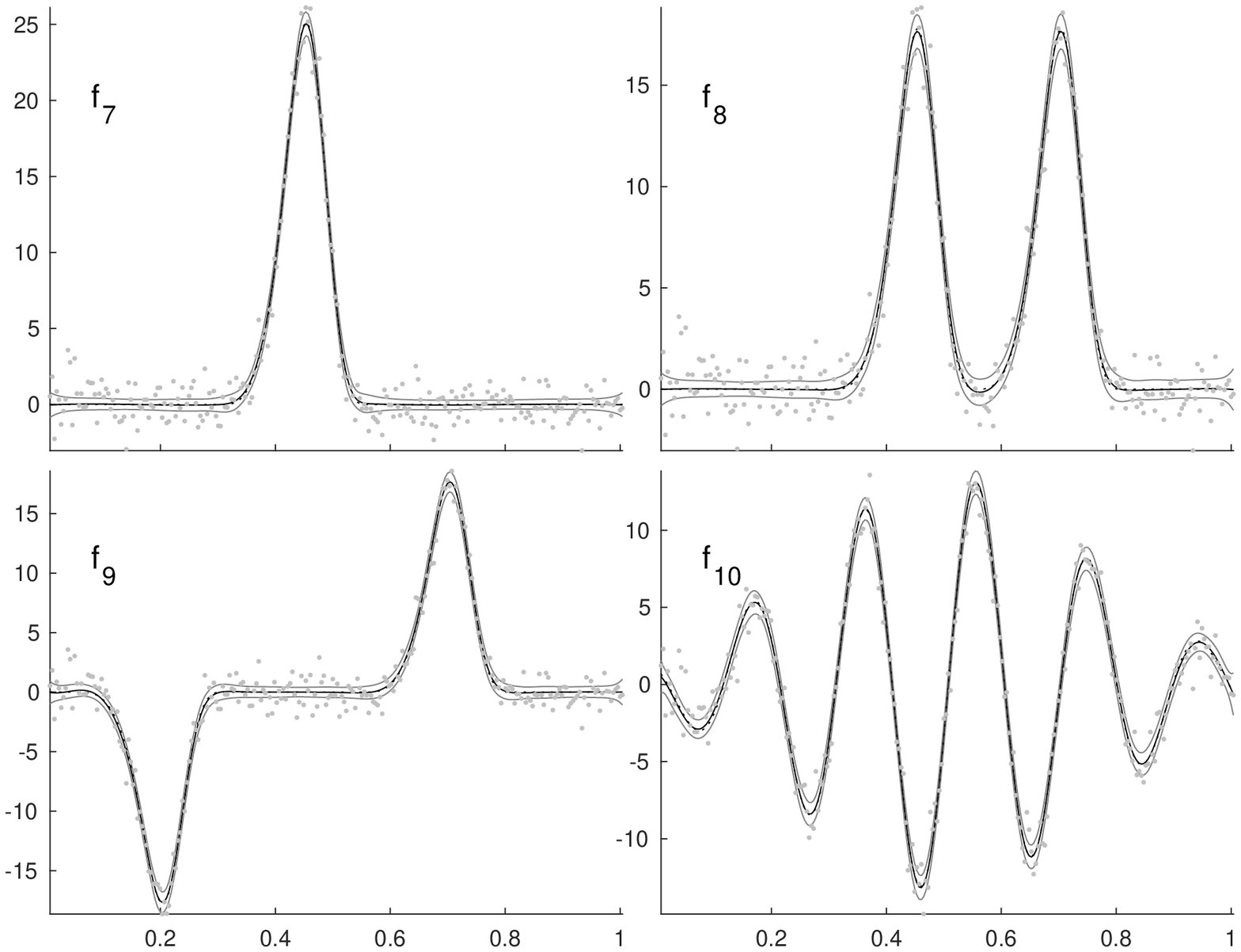}
    \caption{Mean estimated functions (black curve) for benchmark functions $f_7$ to $f_{10}$ at $\snr = 100$.  The true functions are shown as dotted curves
    but they are practically indistinguishable from the mean estimated functions. The gray curves show $\pm 2\sigma$ 
    deviation from the mean function, where $\sigma$ is the estimated 
    standard deviation. The gray dots show an arbitrary 
    data realization for the purpose of visualizing the noise level. 
     The abscissa has the same range for each panel. The algorithm settings used are given by
    the key string ${\rm LP}\_100\_0.1\_100\_{\rm FKM}$, which can be expanded using Table~\ref{tab:labelScheme}.
    %LP_100_0.1_100_FKM
    }
    \label{fig:SNR100_MeanEstSig_7to10}
\end{figure}

With all the design choices fixed, the performance of \fitalgoname can be examined.
This is done in Fig.~\ref{fig:SNR100_MeanEstSig_1to6} and 
Fig.~\ref{fig:SNR100_MeanEstSig_7to10} where the point-wise
sample mean and $\pm 2\sigma$ deviation, $\sigma$ being the sample standard
deviation, are shown for all the benchmark functions. Note that the level of 
noise now is much higher than the examples studied in Sec.~\ref{sec:SNRHigh}.

It is evident from these figures that \fitalgoname is able to
resolve different types of discontinuities as well as  the locations of features such as peaks and sharp 
changes. In interpreting the error envelope, it should be noted that the 
errors at different points are strongly correlated, a fact not reflected in
the point-wise standard deviation. Thus, a typical single estimate
is not an irregular curve bounded by the error envelopes, as would be the case 
for statistically independent point-wise errors, but a smooth function.
Nonetheless, the error envelopes serve to indicate the extent 
to which an estimate can deviate from the true function.

%%%%%%%%%%%%%%%%%%%%%%%
\subsection{Results for $\snr=10$}
\label{sec:SNR10}
\begin{figure}
    \centering
    \includegraphics[scale=0.45]{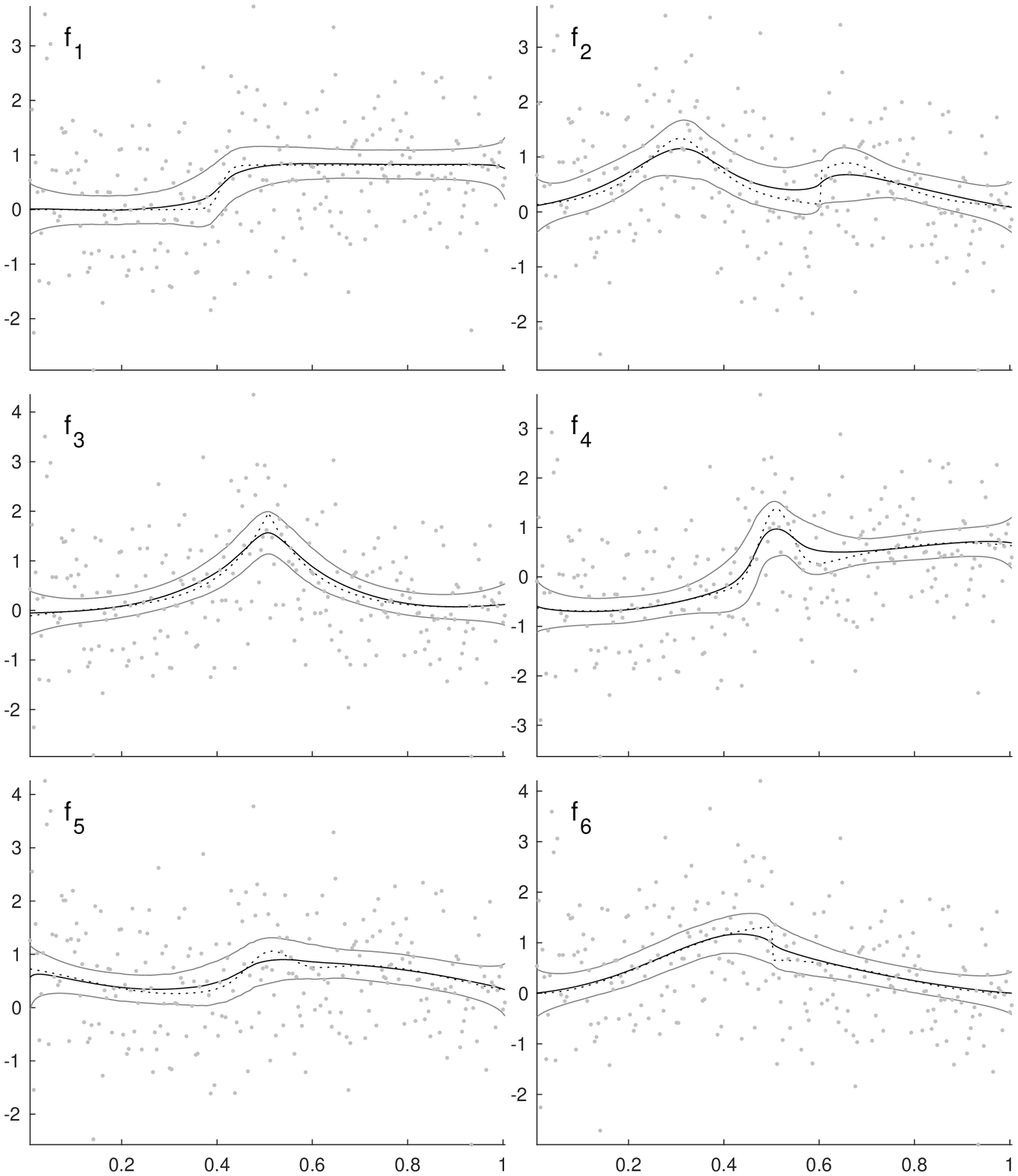}
    \caption{Mean estimated functions (black curve) for benchmark functions $f_1$ to $f_6$ at $\snr = 10$.  The true functions are shown as dotted curves. The gray curves show $\pm 2\sigma$ 
    deviation from the mean function, where $\sigma$ is the estimated 
    standard deviation. The gray dots show an arbitrary 
    data realization for the purpose of visualizing the noise level.
     The abscissa has the same range for each panel. The algorithm settings used are given by
    the key string ${\rm LP}\_10\_5\_50\_{\rm FKM}$, which can be expanded using Table~\ref{tab:labelScheme}.
    %LP_10_5_50_FKM
    }
    \label{fig:SNR10_MeanEstSig_1to6}
\end{figure}

\begin{figure}
    \centering
    \includegraphics[scale=0.45]{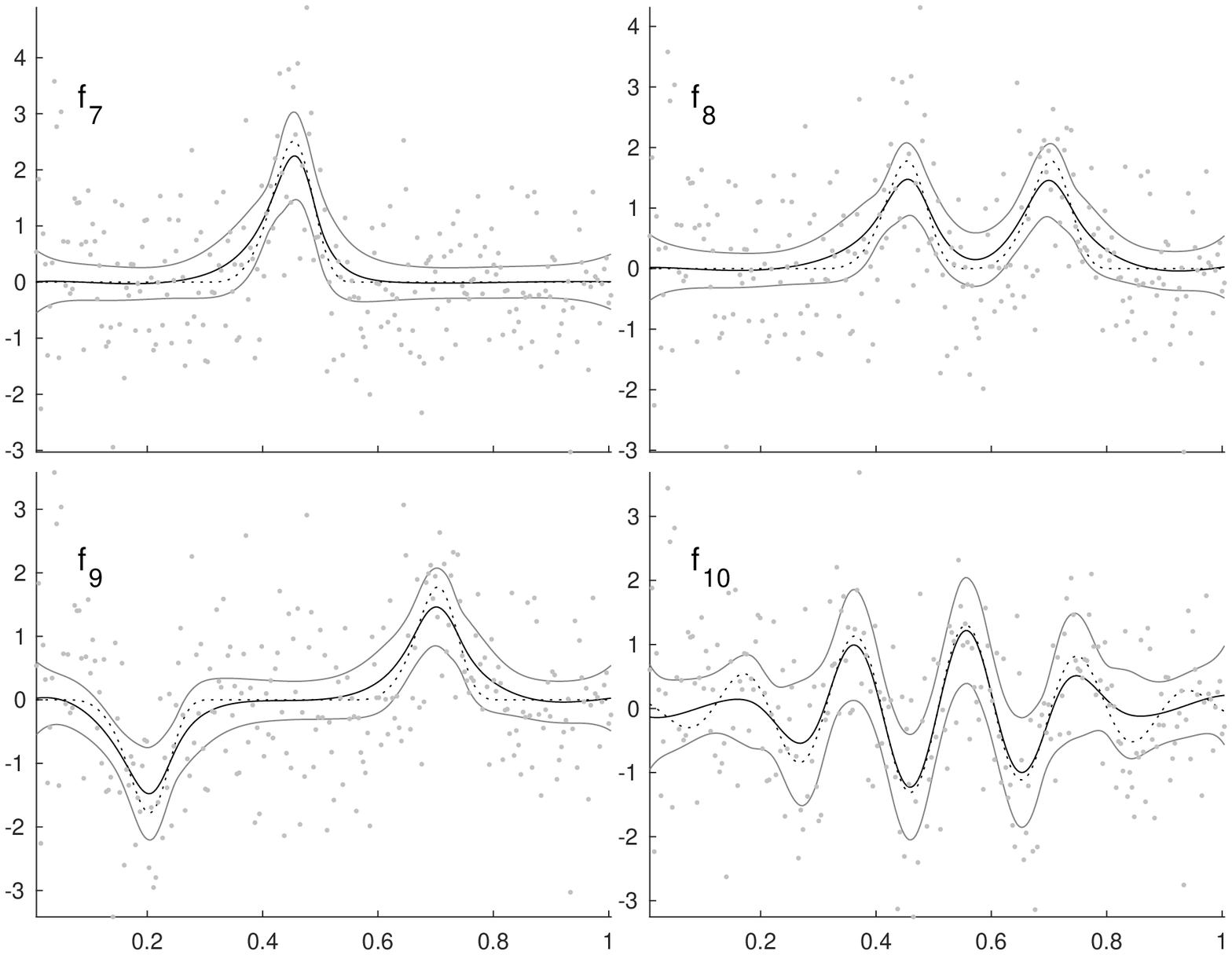}
    \caption{Mean estimated functions (black curve) for benchmark functions $f_7$ to $f_{10}$ at $\snr = 10$.  The true functions are shown as dotted curves. The gray curves show $\pm 2\sigma$ 
    deviation from the mean function, where $\sigma$ is the estimated 
    standard deviation. The gray dots show an arbitrary 
    data realization for the purpose of visualizing the noise level.
     The abscissa has the same range for each panel. The algorithm settings used are given by
    the key string ${\rm LP}\_10\_5\_50\_{\rm FKM}$, which can be expanded using Table~\ref{tab:labelScheme}.
    %LP_10_5_50_FKM
    }
    \label{fig:SNR10_MeanEstSig_7to10}
\end{figure}

Here, we examine the case of high noise level at $\snr=10$. Fig.~\ref{fig:SNR10_MeanEstSig_1to6} and Fig.~\ref{fig:SNR10_MeanEstSig_7to10}
show the point-wise
sample mean and $\pm 2\sigma$ deviation, $\sigma$ being the sample standard
deviation, for all the benchmark functions. The algorithm settings used are
the same as in Sec.~\ref{sec:SNR100} for the $\snr=100$ case 
except for the regulator gain and the 
number of PSO iterations: $\lambda = 5.0$ and $\niter=50$ respectively.

Unlike the $\snr=100$ case, the high noise level masks many of the 
features of the functions. For example, the discontinuities and the non-zero
end values for $f_1$ to $f_6$ are washed out. Thus, the algorithm 
settings to use are not at all as clear cut as before.
In fact, the results presented next
show that alternative settings can show substantial improvements in some cases.

\begin{figure}
    \centering
    \includegraphics[scale=0.45]{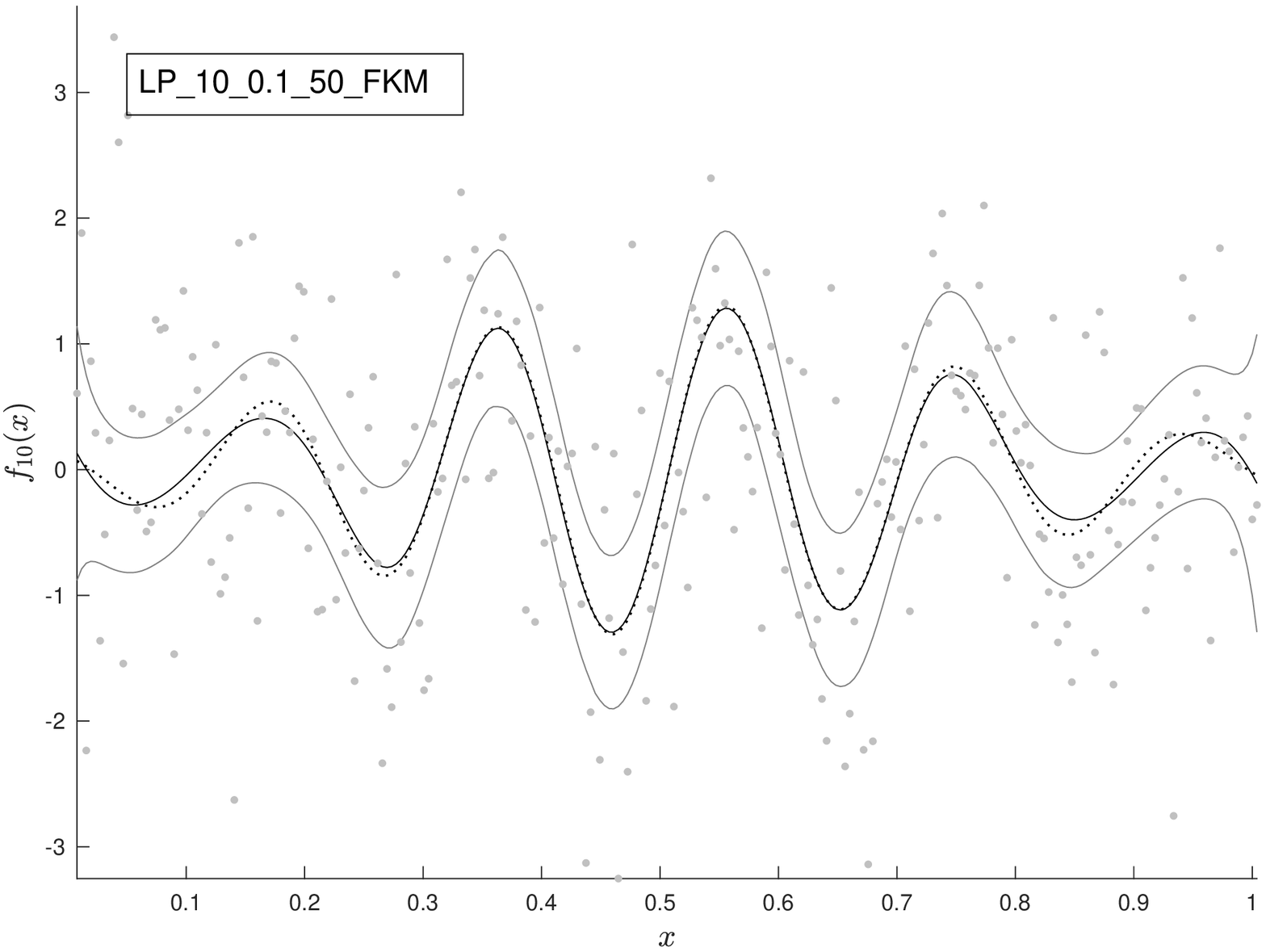}
    \caption{The mean estimated function (black) for benchmark function $f_{10}$ at $\snr = 10$ with regulator gain $\lambda = 0.1$. (The dotted curve shows the true function.) The gray curves show $\pm 2\sigma$ 
    deviation from the mean function, where $\sigma$ is the estimated 
    standard deviation. The gray dots show an arbitrary data realization for the purpose of visualizing the noise level.  The other algorithm settings can be read off from the key shown in the plot legend using Table~\ref{tab:labelScheme}.}
    \label{fig:SNR10_SMBF10_rgain0p1}
\end{figure}

First, as shown in Fig.~\ref{fig:SNR10_SMBF10_rgain0p1}, the estimation of $f_{10}$ actually improves significantly 
when the regulator gain is turned down to $\lambda = 0.1$. While this is the lone 
outlier in the general trend between regulator gain and RMSE (c.f., Fig.~\ref{fig:SNR10_regGainComp}), it points to the importance of choosing
the regulator gain adaptively rather than empirically as done in this paper.  

\begin{figure}[b!]
    \centering
    \includegraphics[scale=0.45]{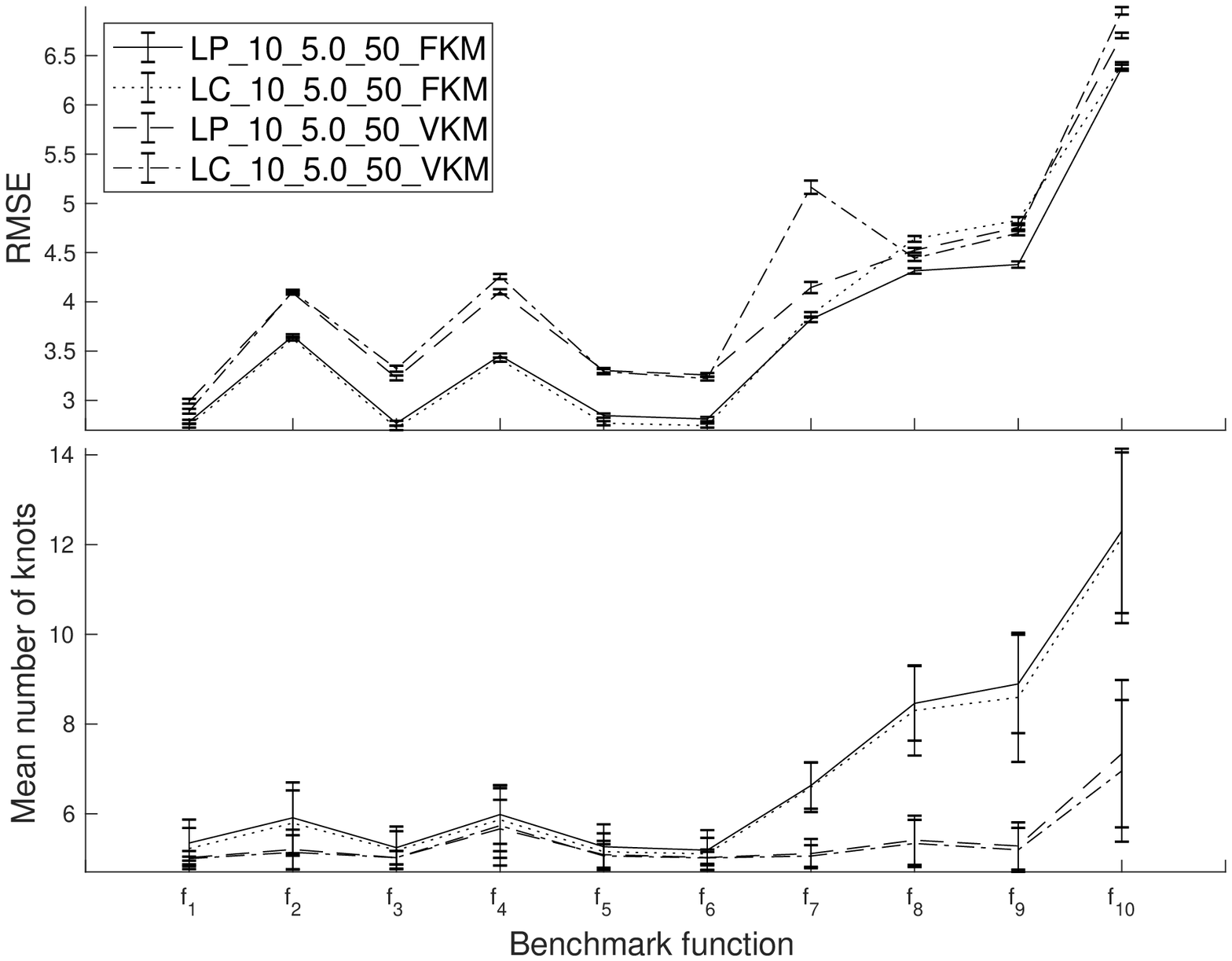}
    \caption{Comparison of plain and centered-monotonic maps 
      under fixed ($F$) and variable ($V$) end knot conditions at $\snr=10$. 
      The top and bottom panels respectively 
      show RMSE and mean number of knots in 
      the best model.
      The other
    algorithm settings used for this plot can be read off  from the key strings 
    shown in the legend using Table~\ref{tab:labelScheme}. The data points correspond to the benchmark functions shown on the abscissa. The error bars show $\pm 1 \sigma$
    deviations, where $\sigma$ is the estimated standard deviation.}
    \label{fig:SNR10_plain_clscheme_fixed_var}
\end{figure}

Next, Fig.~\ref{fig:SNR10_plain_clscheme_fixed_var} examines the effect of 
the knot map and its interplay with fixing the end knots or allowing them to vary. 
Allowing the end knots to vary under either knot map leads to a worse RMSE but the number of knots required in the best fit model is reduced, significantly so for
$f_7$ to $f_{10}$.
A plausible explanation for this is that the high noise level masks the behavior
of the functions at their end points, and freeing up the end knots allows \fitalgoname
to ignore those regions and focus more on the ones where the function value is higher relative to noise.

\begin{figure}
    \centering
    \includegraphics[scale=0.45]{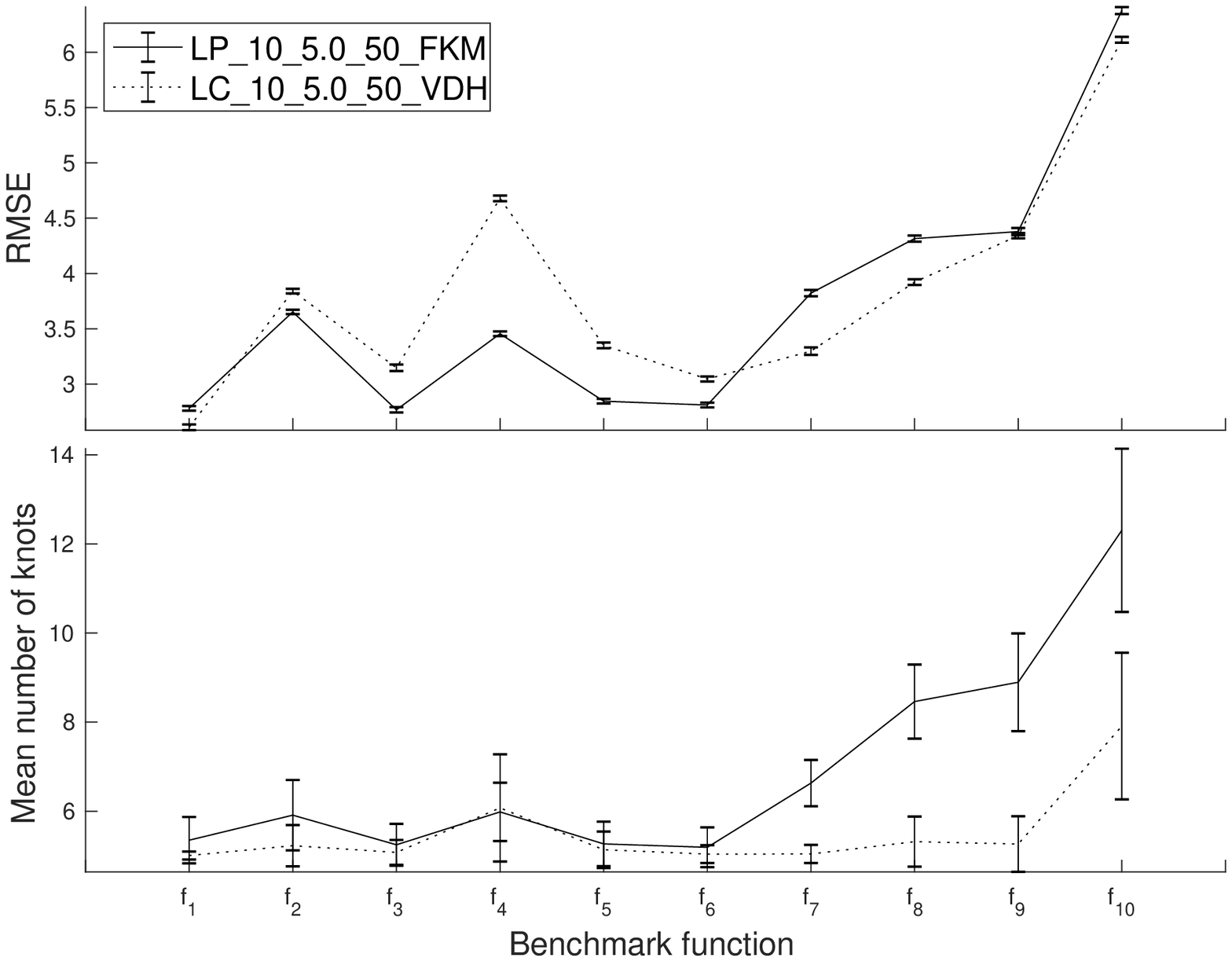}
    \caption{Comparison of plain and centered-monotonic maps
      under the FKM and VDH algorithm settings respectively at $\snr=10$. See
     Table~\ref{tab:labelScheme} for the meaning of these and  other
    algorithm settings given by the key strings in the legend.
    The top and bottom panels respectively 
      show RMSE and mean number of knots in 
      the best model. The data points correspond to the benchmark functions shown on the abscissa. The error bars show $\pm 1 \sigma$
    deviations, where $\sigma$ is the estimated standard deviation.}
    \label{fig:SNR10_plain_clscheme_FKM_VDH}
\end{figure}

Under a given end knot condition, the centered-monotonic map always performs worse
in Fig.~\ref{fig:SNR10_plain_clscheme_fixed_var} albeit the difference is statistically
significant for only a small subset of the benchmark functions.
Remarkably, this behavior is reversed for some of the benchmark functions when
additional changes are made to the design choices. Fig.~\ref{fig:SNR10_plain_clscheme_FKM_VDH} shows the RMSE when the centered-monotonic map and variable end knots are coupled with the 
dropping of end B-splines and healing of knots. Now, the performance is better for functions
$f_7$ to $f_{10}$ relative to the best algorithm settings found from 
Fig.~\ref{fig:SNR10_plain_clscheme_fixed_var}: not only is there a statistically 
significant improvement in the RMSE for these functions but this is achieved with a substantially smaller number of knots. This improvement
comes at the cost, however, 
of significantly worsening the RMSE for the remaining benchmark functions.
%%%%%%%%%%%%%%%%%%%%%%%%%%%%%%%%
\subsection{Effect of bias correction}
\label{sec:comp_bias_correct}
\begin{figure}
    \centering
    \includegraphics[scale=0.45]{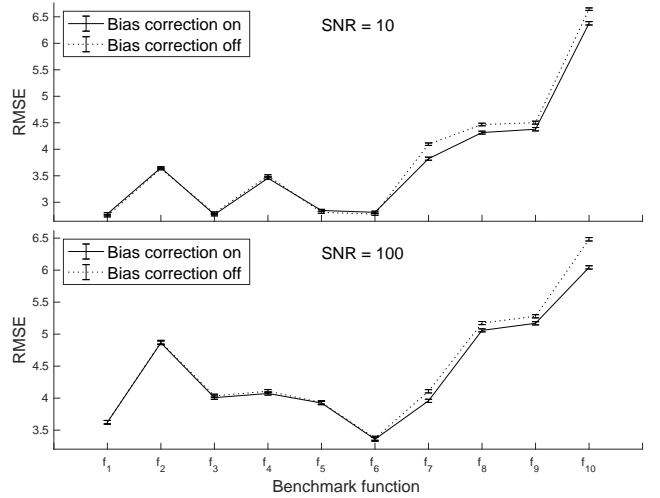}
    \caption{
    The effect of bias correction on root mean squared error (RMSE) for 
     $\snr=10$ (top) and $\snr=100$ (bottom) benchmark functions.
    Solid and dotted curves in each panel correspond to bias correction 
    switched on and off, respectively. The algorithm settings used were 
    ${\rm LP}\_{\snr}\_\lambda\_N_{\rm iter}\_{\rm FKM}$ in all the cases with (top) $\lambda = 5.0$, $\niter=50$, and (bottom) $\lambda = 0.1$, $\niter=100$. These are the fiducial settings used for $\snr=10$ and $\snr=100$ in Sec.~\ref{sec:SNR10} and Sec.~\ref{sec:SNR100}, respectively. The data points correspond to the benchmark functions shown on the abscissa. The error bars show $\pm 1 \sigma$
    deviations, where $\sigma$ is the estimated standard deviation
    }
    \label{fig:biasrmsecomp}
\end{figure}

Fig.~\ref{fig:biasrmsecomp} shows the effect of using the bias correction
step described in Sec.~\ref{sec:bias} on RMSE. We see that bias correction reduces the RMSE for some of the benchmark functions, namely $f_7$ to $f_{10}$, and that the reduction
is more at higher $\snr$ for $f_{10}$. For the remaining benchmark functions, bias correction makes no difference to the
RMSE.  
%%%%%%%%%%%%%%%%
%%%%%%%%%%%%%%%

%%%%%%%%%%%%%%%%%%%%%%%%%%%%%%%%
\subsection{Comparison with other methods}
\label{sec:comp_susr_smsp}
In this section, we compare the performance of \fitalgoname with 
\altalgows~\cite{donoho1995adapting} and smoothing spline~\cite{reinsch1967smoothing,craven1978smoothing} as 
implemented in R~\cite{R-software} (called \altalgors in this paper). The former
is taken from the Matlab-based package \texttt{WaveLab}~\cite{wavelab} and it performs smoothing by thresholding the wavelet coefficients of the given data and applying non-linear shrinkage to threshold-crossing coefficients. For both of these methods, we use default values of their parameters with the following exceptions: for \altalgows we used the ``Hybrid" shrinkage method, while for \altalgors we use GCV to determine the regulator gain. For reproducibility, we list the exact commands used to call these methods:
\begin{itemize}
    \item \texttt{WaveShrink(Y,`Hybrid',L)}: \texttt{Y} is the data to be smoothed and \texttt{L} is the coarsest resolution level of the discrete wavelet transform of \texttt{Y}. 
    \item \texttt{smooth.spline(X,Y,cv=FALSE)}: \texttt{X} is the set of predictor values, \texttt{Y} is the data to be smoothed, and \texttt{cv=FALSE} directs the code to use GCV for regulator gain determination.
\end{itemize}
Each method above was applied to the same dataset as used for producing Fig.~\ref{fig:SNR100_MeanEstSig_1to6}. Statistical summaries, namely, the mean estimated function, the $\pm 2\sigma$ deviation from the mean, and RMSE were obtained following the same procedure as described for \fitalgoname.
A crucial detail: when applying \altalgows,  we use $\texttt{L}\in\{1,2,\ldots,6\}$ and pick the one that gives the lowest RMSE. 

\begin{figure*}
%%%%%%%%%%%%%%%%%%%%%%%%%%%%%%%%%%%%%%%%%%%%%%%%%%%%%%%%%%%
    \centering
    \includegraphics[scale=0.6]{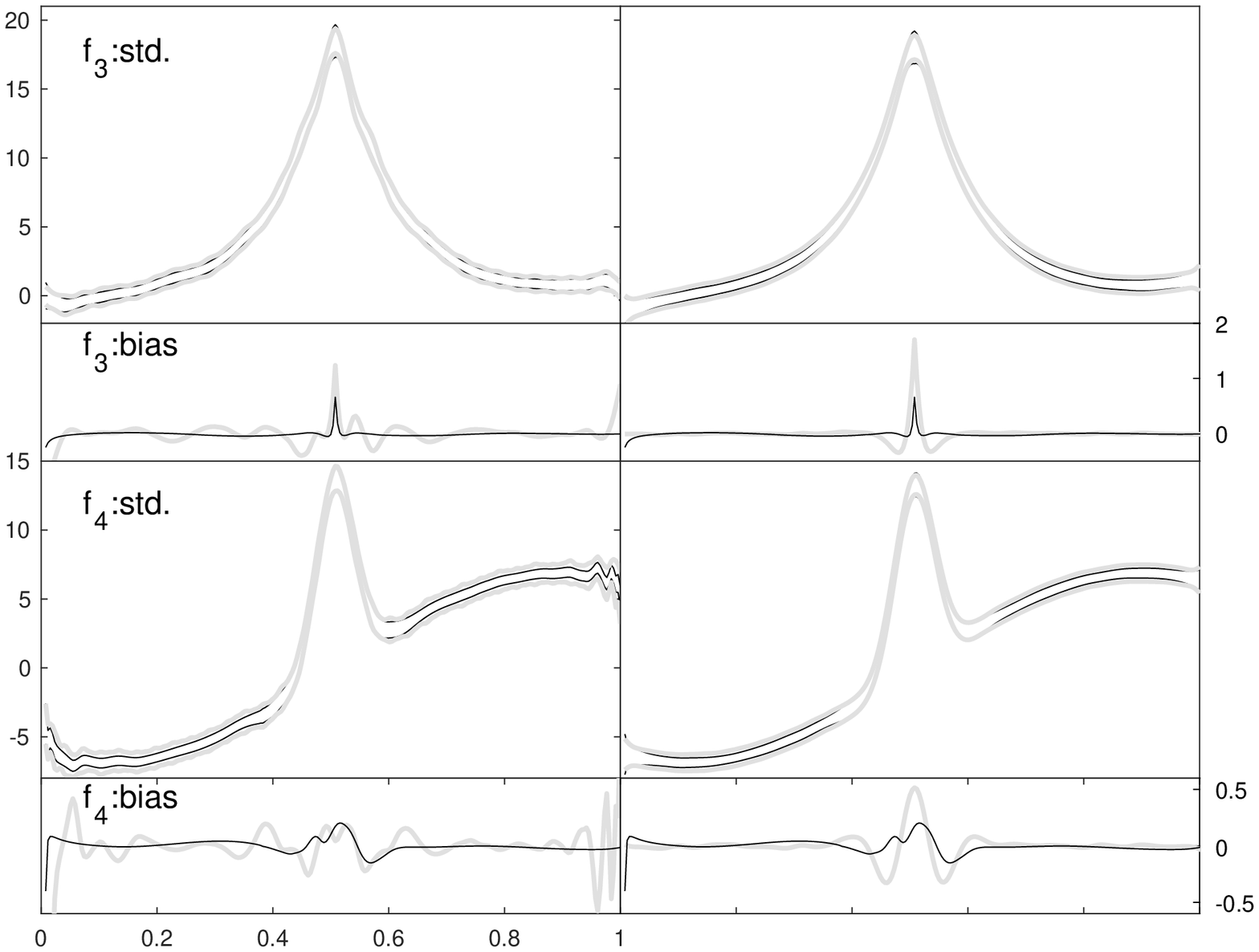}
    \caption{Comparison of \fitalgoname with alternate methods, \altalgows and \altalgors, for benchmark functions $f_3$ and $f_4$ at $\snr = 100$. The left column of figure panels is associated with \altalgows and 
    the right with \altalgors. In each column, (i) the benchmark function used is indicated in each panel, (ii) black curves correspond to \fitalgoname, and (iii) gray curves to the alternate algorithm. For each benchmark function, there are two panels: (i) The one labeled ``std" shows the $\pm 2\sigma$ error envelopes relative to the estimated mean signal from the alternate method;  (ii) the one labeled ``bias" shows the difference between the true function and the estimated mean signal from each method. 
     The abscissa has the same range for each panel. The ordinate values are identical across the panels in a given row. The dataset used and the algorithm settings for \fitalgoname are the same as in Fig.~\ref{fig:SNR100_MeanEstSig_1to6}.
    }
    \label{fig:compPlot_waveshrink_best_worst}
%%%%%%%%%%%%%%%%%COMPSTATS SPRINGER%%%%%%%%%%%%%%%%%%%%%%%%
% \end{figure}
%%%%%%%%%%%%%%%%%%%%%%%%%%%%%%%%%%%%%%%%%%%%%%%%%%%%%%%%%%%
%%%%%%%%%%%%%%%%%%ARXIV PREPRINT%%%%%%%%%%%%%%%%%%%%%%%%%%%
\end{figure*}

\begin{figure*}
%%%%%%%%%%%%%%%%%%%%%%%%%%%%%%%%%%%%%%%%%%%%%%%%%%%%%%%%%%%
    \centering
    \includegraphics[scale=0.6]{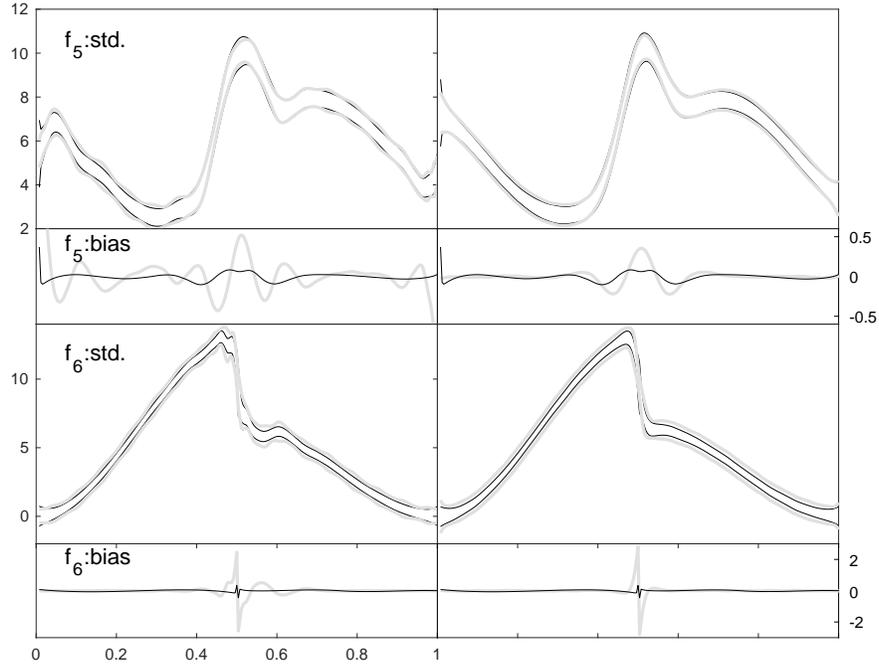}
    \caption{Same as Fig.~\ref{fig:compPlot_waveshrink_best_worst} except for the change in benchmark functions to $f_5$ and $f_6$.
    }
    \label{fig:compPlot_smoothspline_best_worst}
%%%%%%%%%%%%%%%%%COMPSTATS SPRINGER%%%%%%%%%%%%%%%%%%%%%%%%
% \end{figure}
%%%%%%%%%%%%%%%%%%%%%%%%%%%%%%%%%%%%%%%%%%%%%%%%%%%%%%%%%%%
%%%%%%%%%%%%%%%%%%ARXIV PREPRINT%%%%%%%%%%%%%%%%%%%%%%%%%%%
\end{figure*}

The results of the comparison are shown in Table~\ref{tab:SHAPES_WS_RSS_RMSE}, Fig.~\ref{fig:compPlot_waveshrink_best_worst}, and Fig.~\ref{fig:compPlot_smoothspline_best_worst}. Table~\ref{tab:SHAPES_WS_RSS_RMSE}
shows the RMSE values attained by the methods for the 
benchmark functions $f_1$ to $f_6$, all normalized to have ${\rm SNR}=100$. 
Figure~\ref{fig:compPlot_waveshrink_best_worst} shows more details for the 
benchmark functions that produce the best and worst RMSE values for \altalgows. Similarly, Fig.~\ref{fig:compPlot_smoothspline_best_worst} corresponds to
the best and worst benchmark functions for \altalgors. 
%%%%%%%%%%%%%%%%
\begin{table}
    \centering
    \begin{tabular}{|c|c|c|c|}
    \hline
        & \fitalgoname & \altalgows & \altalgors
        \\ \hline
        $f_1$ & $3.62$ &$7.96$ $(4)$ &$4.91$ \\
        $f_2$ & $4.86$ &$7.47$ $(4)$ &$7.39$ \\
        $f_3$ & $4.01$ &$5.82$ $(3)$ &$5.52$ \\
        $f_4$ & $4.07$ &$8.11$ $(4)$ &$5.24$ \\
        $f_5$ & $3.92$ &$6.21$ $(3)$ &$4.19$ \\
        $f_6$ & $3.36$ &$6.93$ $(3)$ &$7.62$ \\
        \hline
    \end{tabular}
    \caption{RMSE values for \fitalgoname, \altalgows, and \altalgors obtained with the same dataset as used for 
    Fig.~\ref{fig:SNR100_MeanEstSig_1to6}. The benchmark functions used are $f_1$
    to $f_6$ at ${\rm SNR}=100$. In the case of \altalgows, the RMSE is the lowest attained over different values of the parameter \texttt{L}. The best value of  \texttt{L} is shown parenthetically.
    }
    \label{tab:SHAPES_WS_RSS_RMSE}
\end{table}
%%%%%%%%%%%%%%%%
%%%%%%%%%%%%%%%%%COMPSTATS SPRINGER%%%%%%%%%%%%%%%%%%%%%%%%
% \begin{figure}
%%%%%%%%%%%%%%%%%%%%%%%%%%%%%%%%%%%%%%%%%%%%%%%%%%%%%%%%%%%
%%%%%%%%%%%%%%%%%%ARXIV PREPRINT%%%%%%%%%%%%%%%%%%%%%%%%%%%
%%%%%%%%%%%%%%%%%%%%%%%%%%%%%%%%%%%%%%%%%%%%%%%%%%%%%%%%%%%
%%%%%%%%%%%%%%%
%%%%%%%%%%%%%%%%%COMPSTATS SPRINGER%%%%%%%%%%%%%%%%%%%%%%%%
% \begin{figure}
%%%%%%%%%%%%%%%%%%%%%%%%%%%%%%%%%%%%%%%%%%%%%%%%%%%%%%%%%%%
%%%%%%%%%%%%%%%%%%ARXIV PREPRINT%%%%%%%%%%%%%%%%%%%%%%%%%%%
%%%%%%%%%%%%%%%%%%%%%%%%%%%%%%%%%%%%%%%%%%%%%%%%%%%%%%%%%%%
%%%%%%%%%%%%%%%

While noted in the caption of Fig.~\ref{fig:compPlot_waveshrink_best_worst}, it is worth re-emphasizing here that the error envelopes of \fitalgoname shown in Fig.~\ref{fig:compPlot_waveshrink_best_worst} and~\ref{fig:compPlot_smoothspline_best_worst} are computed relative to the mean estimated function of the method being compared to, not to that of \fitalgoname itself.  This modification eliminates visual confusion caused by the different biases (i.e., mean estimated functions) of the methods. However, the bias curves shown separately in these figures do use the respective mean estimated function for each method. The actual mean estimated functions and corresponding error envelopes of \fitalgoname  can be seen in Fig.~\ref{fig:SNR100_MeanEstSig_1to6}.

From Table~\ref{tab:SHAPES_WS_RSS_RMSE}, we see that \fitalgoname has 
the lowest RMSE in all cases.   Fig.~\ref{fig:compPlot_waveshrink_best_worst} and Fig.~\ref{fig:compPlot_smoothspline_best_worst} show that this arises from  
 \fitalgoname generally having both a lower  estimation variance (where its error envelope nests within
 that of the other methods)
 as well as a lower bias. Typically, \altalgors has a  lower variance than \fitalgoname around stationary points of the true function but the difference is marginal. In some cases, such as $f_3$ and $f_6$, the bias in the \fitalgoname estimate is significantly lower than that of either \altalgows or \altalgors.  In general,
\altalgows estimates are less smooth than those from either \fitalgoname or \altalgors. This is manifested, for example, in the rougher behavior of the mean estimated function from \altalgows. Both \altalgows and \altalgors
have a much poorer resolution of the jump discontinuity in $f_6$ compared to \fitalgoname (c.f., Fig.~\ref{fig:SNR100_MeanEstSig_1to6}). 
%%%%%%%%%%%%%%%%%%%%%%%%%%%%%%%%%%%%%%%%%%%%%%%%%%%%
\section{Conclusions}
\label{sec:conclusions}
Our results show that the challenge of free knot placement  
in adaptive spline fitting is solvable. The most important element of
the solution is the use of an effective metaheuristic for knot optimization. 
We have shown that lbest PSO is effective in this task. Considering the 
$f_{10}$ benchmark function for example, the best model
found by \fitalgoname reaches
the vicinity of the highest number ($=18$) of  non-repeating knots
considered in this paper.
The good quality of the fit obtained for $f_{10}$ shows that PSO was able to
handle this high-dimensional optimization well. 

Relative to the SNRs used commonly in the literature on adaptive spline fitting, the values of SNR used in this paper, namely ${\rm SNR}=100$ and ${\rm SNR}=10$, can be ranked respectively as being moderate to low.
For the former, discontinuities 
in function values or derivatives were well localized by \fitalgoname 
in all the cases. At the same 
time, the smooth parts of the benchmark functions were also well estimated.  
The estimates from \fitalgoname for low SNR ($=10$) had, naturally, more error. 
In particular, the noise level in all the data realizations was high enough to completely
mask the presence of discontinuities and, thus, they were not well localized.
Nonetheless, even with a conservative error
envelope of $\pm 2 \sigma$ around the mean estimated signal, the overall shape of the
true function is visually clear in all the examples. This shows that the estimated
functions are responding  
to the presence of the true function in the data. 

% Further characterization of the performance of \fitalgoname for the low ${\rm SNR}$
% case requires
% incorporating a hypothesis test, a topic that we are currently investigating. For the 
% case of transient signals represented by $f_7$ to $f_{10}$, the fitness value returned by  \fitalgoname may itself serve as a powerful detection statistic since the
% estimated functions depart quite significantly from the expected behavior
% under the null hypothesis. (Hypotheses testing for $f_7$ was explored in this manner
% in~\cite{mohanty2018swarm} with a method that did not include bias reduction.)
% However, this expectation may need to be modified for the 
% other benchmark functions, in particular for $f_1$. Another issue that comes up in 
% a hypothesis test is that of bias correction: it must be used only if the detection
% threshold of the test is crossed, otherwise estimates under both the null and 
% alternative hypotheses will get amplified. This creates an interesting interplay between
% hypothesis testing and estimation that needs a careful scrutiny.

While we have characterized the performance of \fitalgoname as an estimator in this paper, the observation made above for
the low $\snr$ case suggests that \fitalgoname may also be used to set 
up a hypotheses test. This could be based, say, on the fitness value
returned by \fitalgoname. Note that, being a non-parametric method, \fitalgoname can handle functions with qualitatively disparate behaviors -- from a simple change between two levels  to oscillatory -- 
without requiring any special tuning. Thus, such a hypotheses test would allow the detection
of signals with a wide morphological range. This investigation is in progress.

The dependence of design choices 
on $\snr$, as elucidated in this paper,
does not seem to have been fully appreciated in the literature on 
adaptive spline fitting, probably because the typical scenario considered 
is that of high $\snr$.
While performance of \fitalgoname for $\snr=100$ is found to be 
 fairly robust to the design choices made,
they have a non-negligible affect at $\snr=10$. The nature of the 
true function also influences the appropriate algorithm settings for the latter case.
Fortunately, the settings were found to depend on 
only some coarse features of a function, such as its behavior at
data boundaries ($f_1$ to $f_6$), whether it is transient ($f_7$ to $f_9$),
or whether it is
oscillatory ($f_{10})$. Such features 
are often well-known in a real-world application domain: it is unusual 
to deal with signals that have discontinuities as well 
as signals that are smooth and transient in the same application.
Hence, in most such cases, 
it should be straightforward to pick the best settings for \fitalgoname.

The inclusion of a  penalized spline  regulator 
was critical in \fitalgoname for mitigating the 
problem of knot clustering. For all except one ($f_{10}$)
benchmark functions considered here, the regulator gain was determined empirically
 by simply
 examining a few realizations at each ${\rm SNR}$ with different values of the 
regulator gain $\lambda$.
Ideally, however, $\lambda$
should be determined adaptively from given data using a method such as GCV.  The case of $f_{10}$
at ${\rm SNR}=10$ provides a particularly good test bed in this regard: while  $\lambda = 5.0$ worked well for the other benchmark functions at ${\rm SNR}=10$, the 
RMSE for $f_{10}$ improved significantly when the gain was lowered to $\lambda = 0.1$. Thus, any method for determining $\lambda$ adaptively must be able to handle
this extreme variation. We leave the additional refinement of using
an adaptive regulator gain in \fitalgoname to future work. 

The extension of \fitalgoname to multi-dimensional splines  
and longer data lengths is the next logical step in its development.
It is likely that extending \fitalgoname to
these higher complexity problems will require different PSO variants
than the one used here. 

The codes used in this paper for the FKM option (c.f., Sec.~\ref{sec:SNR100}) are
available in a \texttt{GitHub}
repository at the URL:\\ \texttt{https://github.com/mohanty-sd/SHAPES.git}
%%%%%%%%%%%%%%%%%%%%%%%%%%%%%%%%%%%%%%%%%%%%%%%%%
%%%%%%%%%%%%%%%%%  ARXIV PREPRINT %%%%%%%%%%%%%%%%%%%%%%
\section{Acknowledgements}
%%%%%%%%%%%%%%%%%%%%%%%%%%%%%%%%%%%%%%%%%%%%%%%%%%%%%%%%%

%%%%%%%%%%%%%%%%% COMPTSTATS (SPRINGER) %%%%%%%%%%%%%%%%%
% \begin{acknowledgements}
%%%%%%%%%%%%%%%%%%%%%%%%%%%%%%%%%%%%%%%%%%%%%%%%%%%%%%%%%

The contribution of S.D.M. to this paper was supported by National Science Foundation (NSF) grant PHY-1505861. 
The contribution of E.F. to this paper was supported by NSF grant PHY-1757830.
We acknowledge the Texas Advanced Computing Center (TACC) at The University of Texas at Austin (www.tacc.utexas.edu) for providing HPC resources that have contributed to the research results reported within this paper.
%%%%%%%%%%%%%%%%% COMPTSTATS (SPRINGER) %%%%%%%%%%%%%%%%%
% \end{acknowledgements}
%%%%%%%%%%%%%%%%%%%%%%%%%%%%%%%%%%%%%%%%%%%%%%%%%%%%%%%%%

%%%%%%%%%%%%%%%%%  ARXIV PREPRINT %%%%%%%%%%%%%%%%%%%%%%
 \bibliographystyle{elsarticle-num}

%%%%%%%%%%%%%%%%%%%%%%%%%%%%%%%%%%%%%%%%%%%%%%%%%%%%%%%%%
%%%%%%%%%%%%%%%%% COMPTSTATS (SPRINGER) %%%%%%%%%%%%%%%%%
% \bibliographystyle{spphys}  
%Copy-paste the .bbl file content here
%%%%%%%%%%%%%%%%%%%%%%%%%%%%%%%%%%%%%%%%%%%%%%%%%%%%%%%%
%%%%%%%%%%%%%COMMENT OUT WHEN SUBMITTING%%%%%%%%%%%%%%%
%\bibliography{references.bib,mohanty_bib.bib}

\begin{thebibliography}{10}
\expandafter\ifx\csname url\endcsname\relax
  \def\url#1{\texttt{#1}}\fi
\expandafter\ifx\csname urlprefix\endcsname\relax\def\urlprefix{URL }\fi
\expandafter\ifx\csname href\endcsname\relax
  \def\href#1#2{#2} \def\path#1{#1}\fi

\bibitem{deBoor}
C.~de~Boor, A Practical Guide to Splines (Applied Mathematical Sciences),
  Springer, 2001.

\bibitem{wegman1983splines}
E.~J. Wegman, I.~W. Wright, Splines in statistics, Journal of the American
  Statistical Association 78~(382) (1983) 351--365.

\bibitem{wahba1990spline}
G.~Wahba, Spline models for observational data, Vol.~59, Siam, 1990.

\bibitem{hardle1990applied}
W.~H{\"a}rdle, Applied nonparametric regression, no.~19, Cambridge university
  press, 1990.

\bibitem{wold1974spline}
S.~Wold, Spline functions in data analysis, Technometrics 16~(1) (1974) 1--11.

\bibitem{burchard1974splines}
H.~G. Burchard, Splines (with optimal knots) are better, Applicable Analysis
  3~(4) (1974) 309--319.

\bibitem{jupp1978approximation}
D.~L. Jupp, Approximation to data by splines with free knots, SIAM Journal on
  Numerical Analysis 15~(2) (1978) 328--343.

\bibitem{luo1997hybrid}
Z.~Luo, G.~Wahba, Hybrid adaptive splines, Journal of the American Statistical
  Association 92~(437) (1997) 107--116.

\bibitem{smith1982curve}
P.~L. Smith, Curve fitting and modeling with splines using statistical variable
  selection techniques, Tech. rep., NASA, Langley Research Center, Hampton, VA,
  report NASA,166034 (1982).

\bibitem{lyche1988data}
T.~Lyche, K.~M{\o}rken, A data-reduction strategy for splines with applications
  to the approximation of functions and data, IMA Journal of Numerical analysis
  8~(2) (1988) 185--208.

\bibitem{friedman1989flexible}
J.~H. Friedman, B.~W. Silverman, Flexible parsimonious smoothing and additive
  modeling, Technometrics 31~(1) (1989) 3--21.

\bibitem{friedman1991}
J.~H. Friedman, Multivariate adaptive regression splines, The Annals of
  Statistics 19~(1) (1991) 1--67.

\bibitem{stone1997polynomial}
C.~J. Stone, M.~H. Hansen, C.~Kooperberg, Y.~K. Truong, et~al., Polynomial
  splines and their tensor products in extended linear modeling: 1994 wald
  memorial lecture, The Annals of Statistics 25~(4) (1997) 1371--1470.

\bibitem{golub1979generalized}
G.~H. Golub, M.~Heath, G.~Wahba, Generalized cross-validation as a method for
  choosing a good ridge parameter, Technometrics 21~(2) (1979) 215--223.

\bibitem{zhou2001spatially}
S.~Zhou, X.~Shen, Spatially adaptive regression splines and accurate knot
  selection schemes, Journal of the American Statistical Association 96~(453)
  (2001) 247--259.

\bibitem{park2007b}
H.~Park, J.-H. Lee, B-spline curve fitting based on adaptive curve refinement
  using dominant points, Computer-Aided Design 39~(6) (2007) 439--451.

\bibitem{kang2015knot}
H.~Kang, F.~Chen, Y.~Li, J.~Deng, Z.~Yang, Knot calculation for spline fitting
  via sparse optimization, Computer-Aided Design 58 (2015) 179--188.

\bibitem{luo2019knot}
J.~Luo, H.~Kang, Z.~Yang, Knot calculation for spline fitting based on the
  unimodality property, Computer Aided Geometric Design 73 (2019) 54--69.

\bibitem{geneticalgo}
M.~Mitchell, An Introduction to Genetic Algorithms by Melanie Mitchell,
  Bradford, 1998.

\bibitem{ulker2009automatic}
E.~{\"U}lker, A.~Arslan, Automatic knot adjustment using an artificial immune
  system for b-spline curve approximation, Information Sciences 179~(10) (2009)
  1483--1494.

\bibitem{green1995reversible}
P.~J. Green, Reversible jump markov chain monte carlo computation and bayesian
  model determination, Biometrika 82~(4) (1995) 711--732.

\bibitem{pittman2002adaptive}
J.~Pittman, Adaptive splines and genetic algorithms, Journal of Computational
  and Graphical Statistics 11~(3) (2002) 615--638.

\bibitem{dimatteo2001bayesian}
I.~DiMatteo, C.~R. Genovese, R.~E. Kass, Bayesian curve-fitting with free-knot
  splines, Biometrika 88~(4) (2001) 1055--1071.

\bibitem{YOSHIMOTO2003751}
F.~Yoshimoto, T.~Harada, Y.~Yoshimoto, Data fitting with a spline using a
  real-coded genetic algorithm, Computer-Aided Design 35~(8) (2003) 751 -- 760.

\bibitem{miyata2003adaptive}
S.~Miyata, X.~Shen, Adaptive free-knot splines, Journal of Computational and
  Graphical Statistics 12~(1) (2003) 197--213.

\bibitem{galvez2011efficient}
A.~G{\'a}lvez, A.~Iglesias, Efficient particle swarm optimization approach for
  data fitting with free knot b-splines, Computer-Aided Design 43~(12) (2011)
  1683--1692.

\bibitem{mohanty2012particle}
S.~D. Mohanty, Particle swarm optimization and regression analysis {I},
  Astronomical Review 7~(2) (2012) 29--35.

\bibitem{PSO}
J.~Kennedy, R.~C. Eberhart, Particle swarm optimization, in: Proceedings of the
  IEEE International Conference on Neural Networks: Perth, WA, Australia,
  Vol.~4, IEEE, 1995, p. 1942.

\bibitem{SEECR-PhysRevD.96.102008}
S.~D. Mohanty, Spline based search method for unmodeled transient gravitational
  wave chirps, Physical Review D 96 (2017) 102008.
\newblock \href {http://dx.doi.org/10.1103/PhysRevD.96.102008}
  {\path{doi:10.1103/PhysRevD.96.102008}}.

\bibitem{mohanty2018swarm}
S.~D. Mohanty, Swarm Intelligence Methods for Statistical Regression, Chapman
  and Hall/CRC, 2018.

\bibitem{ruppert2003semiparametric}
D.~Ruppert, M.~P. Wand, R.~J. Carroll, Semiparametric regression, Vol.~12,
  Cambridge University Press, 2003.

\bibitem{reinsch1967smoothing}
C.~H. Reinsch, Smoothing by spline functions, Numerische mathematik 10~(3)
  (1967) 177--183.

\bibitem{craven1978smoothing}
P.~Craven, G.~Wahba, Smoothing noisy data with spline functions, Numerische
  mathematik 31~(4) (1978) 377--403.

\bibitem{R-software}
{R Core Team}, {R: A Language and Environment
  for Statistical Computing}, R Foundation for Statistical Computing, Vienna,
  Austria (2019).
\newline\urlprefix\url{https://www.R-project.org/}

\bibitem{wahba_donoho2002wavelet}
G.~Wahba, {\it In discussion} {Wavelet} shrinkage: {Asymptopia?} with
  discussion and a reply by {Donoho, D. L.}, {Johnstone, I. M.},
  {Kerkyacharian, G.} and {Picard, D.}, Journal of the Royal Statistical
  Society Series B 57 (2002) 545--564.

\bibitem{storlie2010locally}
C.~B. Storlie, H.~D. Bondell, B.~J. Reich, A locally adaptive penalty for
  estimation of functions with varying roughness, Journal of Computational and
  Graphical Statistics 19~(3) (2010) 569--589.

\bibitem{liu2010data}
Z.~Liu, W.~Guo, Data driven adaptive spline smoothing, Statistica Sinica (2010)
  1143--1163.

\bibitem{wang2013smoothing}
X.~Wang, P.~Du, J.~Shen, Smoothing splines with varying smoothing parameter,
  Biometrika 100~(4) (2013) 955--970.

\bibitem{curry1947spline}
H.~B. Curry, I.~J. Schoenberg, On spline distributions and their limits-the
  polya distribution functions, Bulletin of the American Mathematical Society
  53~(11) (1947) 1114--1114.

\bibitem{wand2000comparison}
M.~P. Wand, A comparison of regression spline smoothing procedures,
  Computational Statistics 15~(4) (2000) 443--462.

\bibitem{claeskens2009asymptotic}
G.~Claeskens, T.~Krivobokova, J.~D. Opsomer, Asymptotic properties of penalized
  spline estimators, Biometrika 96~(3) (2009) 529--544.

\bibitem{eilers1996flexible}
P.~H. Eilers, B.~D. Marx, Flexible smoothing with b-splines and penalties,
  Statistical science (1996) 89--102.

\bibitem{eilers2015twenty}
P.~H. Eilers, B.~D. Marx, M.~Durb{\'a}n, Twenty years of p-splines, SORT:
  statistics and operations research transactions 39~(2) (2015) 0149--186.

\bibitem{ruppert2000theory}
D.~Ruppert, R.~J. Carroll, Theory \& methods: Spatially-adaptive penalties for
  spline fitting, Australian \& New Zealand Journal of Statistics 42~(2) (2000)
  205--223.

\bibitem{krivobokova2008fast}
T.~Krivobokova, C.~M. Crainiceanu, G.~Kauermann, Fast adaptive penalized
  splines, Journal of Computational and Graphical Statistics 17~(1) (2008)
  1--20.

\bibitem{scheipl2009locally}
F.~Scheipl, T.~Kneib, Locally adaptive bayesian p-splines with a
  normal-exponential-gamma prior, Computational Statistics \& Data Analysis
  53~(10) (2009) 3533--3552.

\bibitem{yang2017adaptive}
L.~Yang, Y.~Hong, Adaptive penalized splines for data smoothing, Computational
  Statistics \& Data Analysis 108 (2017) 70--83.

\bibitem{krivobokova2013smoothing}
T.~Krivobokova, Smoothing parameter selection in two frameworks for penalized
  splines, Journal of the Royal Statistical Society: Series B (Statistical
  Methodology) 75~(4) (2013) 725--741.

\bibitem{ruppert2002selecting}
D.~Ruppert, Selecting the number of knots for penalized splines, Journal of
  computational and graphical statistics 11~(4) (2002) 735--757.

\bibitem{DEBOOR197250}
C.~de~Boor, On calculating with b-splines, Journal of Approximation Theory
  6~(1) (1972) 50 -- 62.

\bibitem{ramsay_silverman_functional}
J.~Ramsay, B.~W. Silverman, Functional data analysis, Springer series in
  statistics, Springer-Verlag New York, 1997.

\bibitem{lindstrom1999penalized}
M.~J. Lindstrom, Penalized estimation of free-knot splines, Journal of
  Computational and Graphical Statistics 8~(2) (1999) 333--352.

\bibitem{goepp2018spline}
V.~Goepp, O.~Bouaziz, G.~Nuel, Spline regression with automatic knot selection,
  arXiv preprint arXiv:1808.01770.

\bibitem{akaike1998information}
H.~Akaike, Information theory and an extension of the maximum likelihood
  principle, in: Selected Papers of Hirotugu Akaike, Springer, 1998, pp.
  199--213.

\bibitem{engelbrecht2005fundamentals}
A.~P. Engelbrecht, Fundamentals of computational swarm intelligence, Vol.~1,
  Wiley Chichester, 2005.

\bibitem{lbestTopology}
J.~Kennedy, Small worlds and mega-minds: effects of neighborhood topology on
  particle swarm performance, in: Proceedings of the 1999 Congress on
  Evolutionary Computation-CEC99 (Cat. No. 99TH8406), Vol.~3, IEEE, 1999, pp.
  1931--1938.

\bibitem{normandin2018particle}
M.~E. Normandin, S.~D. Mohanty, T.~S. Weerathunga, Particle swarm optimization
  based search for gravitational waves from compact binary coalescences:
  Performance improvements, Physical Review D 98~(4) (2018) 044029.

\bibitem{solis1981minimization}
F.~J. Solis, R.~J.-B. Wets, Minimization by random search techniques,
  Mathematics of Operations Research 6~(1) (1981) 19--30.

\bibitem{bratton2007defining}
D.~Bratton, J.~Kennedy, Defining a standard for particle swarm optimization,
  in: Swarm Intelligence Symposium, 2007. SIS 2007. IEEE, IEEE, 2007, pp.
  120--127.

\bibitem{calvin-siuro}
C.~Leung, Estimation of unmodeled gravitational wave transients with spline
  regression and particle swarm optimization, SIAM Undergraduate Research
  Online (SIURO) 8.

\bibitem{denison1998automatic}
D.~Denison, B.~Mallick, A.~Smith, Automatic bayesian curve fitting, Journal of
  the Royal Statistical Society: Series B (Statistical Methodology) 60~(2)
  (1998) 333--350.

\bibitem{li2006bayesian}
M.~Li, Y.~Yan, Bayesian adaptive penalized splines, Journal of Academy of
  Business and Economics 2 (2006) 129--141.

\bibitem{lee2002algorithms}
T.~Lee, On algorithms for ordinary least squares regression spline fitting: a
  comparative study, Journal of statistical computation and simulation 72~(8)
  (2002) 647--663.

\bibitem{matlab}
{Matlab Release2018b}, {The} {MathWorks}, {Inc.}, {Natick}, {Massachusetts},
  {United} {States}.; {\tt http://www.mathworks.com/}.

\bibitem{donoho1995adapting}
D.~L. Donoho, I.~M. Johnstone, Adapting to unknown smoothness via wavelet
  shrinkage, Journal of the american statistical association 90~(432) (1995)
  1200--1224.

\bibitem{wavelab}
X.~Huo, M.~Duncan, O.~Levi, J.~Buckheit, S.~Chen, D.~Donoho, I.~Johnstone,
  \href{http://statweb.stanford.edu/{\textasciitilde{}}wavelab/Wavelab\_850/AboutWaveLab.pdf}{About
  wavelab}.
\newline\urlprefix\url{http://statweb.stanford.edu/~wavelab/Wavelab_850/AboutWaveLab.pdf}

\end{thebibliography}
%%%%%%%%%%%%%%%%%%%%%%%%%%%%%%%%%%%%%%%%%%%%%%%%%%%%%%%
\end{document}